# Ultrafast evanescent heat transfer across solid interfaces via hyperbolic phonon-polaritons in hexagonal boron nitride

William Hutchins,[1] John A. Tomko,[1] Dan M. Hirt,[1] Saman Zare,[1] Joseph R. Matson,[2] Katja Diaz-Granados,[2] Mingze He,[3] Thomas Pfeifer,[1] Jiahan Li,[4] James Edgar,[4] Jon-Paul Maria,[5] Joshua D. Caldwell,[2,3,&] Patrick E. Hopkins[1,6,7,%]

1. Department of Mechanical and Aerospace Engineering, University of Virginia, Charlottesville, VA 22904, USA
2. Interdisciplinary Materials Science Program, Vanderbilt University, Nashville, TN 37212, USA
3. Department of Mechanical Engineering, Vanderbilt University, Nashville, TN 37212, USA
4. Tim Taylor Dept. of Chemical Engineering, Kansas State University, Manhattan, KS, USA.
5. Department of Materials Science and Engineering, Pennsylvania State University, University Park, Pennsylvania 16802, USA
6. Department of Materials Science and Engineering, University of Virginia, Charlottesville, VA 22904, USA
7. Department of Physics, University of Virginia, Charlottesville, VA 22904, USA

& Corresponding Author: josh.caldwell@vanderbilt.edu
% Corresponding Author: phopkins@virginia.edu

The efficiency of phonon-mediated heat transport is limited by the intrinsic atomistic properties of materials, seemingly providing an upper limit to heat transfer in materials and across their interfaces. The typical speeds of conductive transport, which are inherently limited by the chemical bonds and atomic masses, dictate how quickly heat will move in solids. Given that phonon-polaritons (PhPs), or coupled phonon-photon modes, can propagate at speeds approaching 1% of the speed of light – orders of magnitude faster than transport within a pure diffusive phonon conductor – we demonstrate that volume-confined, hyperbolic phonon-polariton (HPhP) modes supported by many biaxial polar crystals can couple energy across solid-solid interfaces at an order of magnitude higher rates than phonon-phonon conduction alone. Using pump-probe thermoreflectance with a mid-infrared, tunable, probe pulse with sub-picosecond resolution, we demonstrate remote and spectrally selective excitation of the HPhP modes in hexagonal boron nitride (hBN) in response to radiative heating from a thermally emitting gold source. Our work demonstrates a new avenue for interfacial heat transfer based on broadband radiative coupling from a hot spot in a gold film to hBN HPhPs, independent of the broad spectral mismatch between the pump (visible) and probe (mid-IR) pulses employed. This methodology can be used to bypass the intrinsically limiting phonon-phonon conductive pathway, thus providing an alternative means of heat transfer across interfaces. Further, our time-resolved measurements of the temperature changes of the HPhP modes in hBN show that through polaritonic coupling, a material can transfer heat across and away from an interface at rates orders of magnitude faster than diffusive phonon speeds intrinsic to the material, thus demonstrating a pronounced thermal transport enhancement in hBN via phonon-polariton coupling.

Across nearly all heterogeneous dielectric interfaces, heat is dissipated via conductive processes driven by acoustic phonons, or other vibrational interactions, which limit thermal boundary conductance[1–3]. Bose-Einstein statistics of nearly thermalized phononic distributions dictate that the typical vibrational energies that contribute to heat transfer across and away from interfaces are lower-energy, higher-group-velocity acoustic phonons, which, at room temperature, are on the order of a few to 10 THz. Due to their larger heat capacities[4], optical phonon branches play an important role in mediating thermal transport[5]. However, the intrinsically slow group velocities[6] restrict these modes' abilities to spread heat away from interfaces and hot spots, thereby putting the burden of thermal dissipation back on the acoustic modes.

In this work, we examine an alternative mechanism to transfer energy across and away from all-solid heterogeneous interfaces by harnessing the local evanescent fields resulting from a hot radiating source to directly launch volume-confined, hyperbolic phonon polaritons (HPhPs) within the frequency range bound by the transverse (TO) and longitudinal (LO) optical phonons of a polar crystalline material, which is referred to as the Reststrahlen band[7]. Hyperbolic polaritons are supported within spectral ranges where a material or structure exhibits dielectric permittivity that are opposite in sign along different crystal axes[8]. Originally explored in metal/dielectric superlattices in the form of hyperbolic metamaterials, HPhPs have been of significant research interest as they can be supported in a homogeneous, low-loss film, with hexagonal boron nitride (hBN) serving as an exemplary material in this regard[9–11]. Isotopic enrichment can also be leveraged to tune the Reststrahlen band and significantly increase the phonon lifetimes (and thus, polariton propagation lengths) of hBN and therefore tune the efficiency of the coupling mechanism[10,12,13], which here was employed to provide optimal propagation lengths. More recently, it has been demonstrated that natural, low-symmetry (highly anisotropic) crystals such as orthorhombic $MoO_3$[14,15] and $V_2O_5$ [16], and monoclinic crystals such as β-$Ga_2O_3$[17] and $CdWO_4$[18] can support HPhPs that can be restricted to propagate along a specific direction in space, with the latter two supporting so-called hyperbolic shear polaritons that are not only highly directional, but also exhibit a propagation direction that is frequency-dependent. As such, if these modes can be induced to carry heat, similar studies could also be widely applied to other hyperbolic materials along specific crystallographic directions via optimal material selection. Specifically, $MoO_3$[15] and twisted metastructures[19] also support highly directional confined HPhP modes, and heating of such modes would allow for directional thermal mitigation of hot spots across interfaces.

While optical phonons typically cannot be used to efficiently conduct heat away from interfaces due to their slow group velocities, this seemingly insurmountable intrinsic material limitation can be overcome by transducing the broadband radiative thermal energy from a thin metal film in the near field into the HPhPs supported within the underlying anisotropic medium, passing this energy across the heterogeneous interface. Previously, HPhPs in hBN have been demonstrated to have propagation lengths of several micrometers,[20,21] enabling applications such as hyperlensing[22–24] and chemical sensing, to enhance thermal transport beyond phonon-limited processes,[25] and a Reststrahlen band that can exhibit reasonable emission even at near room temperatures.[26] Literature has even suggested that this process can occur at interfaces between 2D van der Waal materials,[27–29] warranting the study of this process at 3D contacts, a direct experimental measurement of which would demonstrate the broad applicability of this unique heat transfer process. Through exploiting time-resolved infrared pump-probe measurements, we directly measure the picosecond cooling of indirectly heated optical modes[30,31] in a nearly isotopically pure hexagonal boron nitride ($h^{11}BN$; >99%) flake after the optical pumping of an adjacent gold contact, thereby demonstrating a mechanism for interfacial cooling that is orders of magnitude faster than typical phonon-mediated processes. Previous works have demonstrated fast, efficient near field radiative heat transfer (NFRHT) across narrow vacuum gaps between metallic and dielectric plates,[26] but this approach relies on the inclusion of a nm-scale vacuum gap, thus limiting its utility in certain practices. Our work proves that thermal boundary conductance across heterogeneous interfaces and thermal dissipation in a material away from interfaces can be significantly enhanced through the coupling of a broadband radiating hot spot into propagating PhPs, a finding that surpasses intrinsic material limitations of thermal coupling at interfaces. In doing so, we establish a new paradigm for interfacial thermal transport that moves beyond the nearly century-old concepts and theories of phonon coupling at dielectric interfaces originally discovered by Kapitza.[32]

Enhanced thermal dissipation in solids via polaritonic heat transfer has been theoretically predicted[33] and the prospect of enhanced speeds and even directional control of heat transfer is enabled by the exotic properties of PhPs, such as long propagation lengths,[20,21] anisotropic and nonuniform control of their propagation[17,34–36] and speeds that approach percent levels of the speed of light.[37–40] Experimental works have shown the ability of HPhPs to enhance thermal conductance in dielectrics,[25,41]. However, experimental evidence differentiating the coupled PhP modes from the thermalized phonon modes intrinsic to the material is still lacking. Thus, the first major finding of our work is the ability to both spectrally and temporally resolve the thermally excited HPhP modes in hBN, providing clear evidence of the more efficient thermal transport that these hot HPhP carriers can provide.

Secondly, our results illustrate that HPhP thermal coupling can occur across an all-solid heterogenous interface, and further, that the HPhP modes do not need to be coherently excited, but rather can be optically stimulated by a graybody radiating in the near field. The latter observation is consistent with the seminal work of Greffet et al.,[42] where thermal radiation from a heated SiC diffraction grating resulted in the excitation of propagating SPhPs, which once stimulated were outcoupled via diffraction, giving rise to spatially coherent IR radiation. It is well known that polaritonic modes can enhance heat transfer via NFRHT processes between two radiating bodies separated by a vacuum gap that is less than the Wien wavelength.[43–48] Our work unlocks the possibility of utilizing this NFRHT enhancement at solid-solid interfaces in intimate atomistic contact, by showing that this broadband radiative cooling effect is driven by PhP coupling that can delocalize the thermal energy from a radiating spot into an adjacent hyperbolic medium. This finding will redefine cooling of hot spots in materials and systems limited or dictated by interfacial resistances, such as high power or high frequency electronics,[49] photonic circuits,[50] memory,[51] and thermophotovoltaics.[52]

Our sample geometry is designed to provide a direct observation of spectrally dependent radiative thermal coupling from a thermally excited source (50-nm Au pad) into HPhPs supported in h$^{11}$BN, referred to hence forth at hBN (Fig. 1a). In our pump-probe experiments, we excite the Au pad (Fig. 1b) with a 520-nm pump pulse (incident fluence of 300 MW m$^{-2}$) and a 200-μm 1/e$^2$ diameter probe pulse centered on the Au pad. The central Au pad acts as the focal point of our experiment. We first overlap our pump and probe on the Si substrate. Then, we alter the position of the sample such that the green pump is focused on the circular pad. The rectangular pads serve as reflectors for the polaritons. This configuration creates standing polaritons, similar to Chen et al.[19] in the space spanned by the probing region, which should increase our pump-probe signal. The different distances were used in order to be agnostic to the mean free paths of different HPhP modes. The thermoreflectance changes, measured by the reflected probe pulses from the hBN as a function of time after pump absorption by the Au, are monitored as a function of probe wavelength, which are tuned from below to above the Reststrahlen band of hBN (~1355 – 1610 cm$^{-1}$).[12] Immediately following the pump absorption on the surface of the Au, the Au pad is rapidly volumetrically thermalized via ballistic electron transport,[53] which despite the low thermal emissivity of metals within the mid-IR will radiate efficiently in the near field due to the large imaginary contribution to the Au dielectric permittivity (e.g., optical loss). Such near-field radiation is composed of high-momentum, evanescent fields that can directly match the momenta of the HPhPs within the hBN slab. This coupling mechanism is present through the entire area of the Au/hBN interface, a mechanism that our experiment is designed to detect with our tunable IR probe energies. The enhanced coupling between the optical modes and the radiation underneath the pad give rise to large thermoreflectance optical signals within the Reststrahlen band and at the TO phonon frequency of hBN, as shown in the contour plots of the thermoreflectance spectra vs. Pump-probe delay time in Fig. 1c. In contrast, in the absence of the Au pad, we do not observe any such increase in thermoreflectance signal, as the hBN is transparent to the pump wavelength. This large spectral dependence on our thermoreflectance signal clearly illustrates that the underlying Si substrate is not playing any significant role, as even though it will absorb the pump photons, it does not induce a significant transient spectral response, further demonstrated by supplemental spectral measurements on identical Si substrates (see Fig. S5 in Supplementary Information).

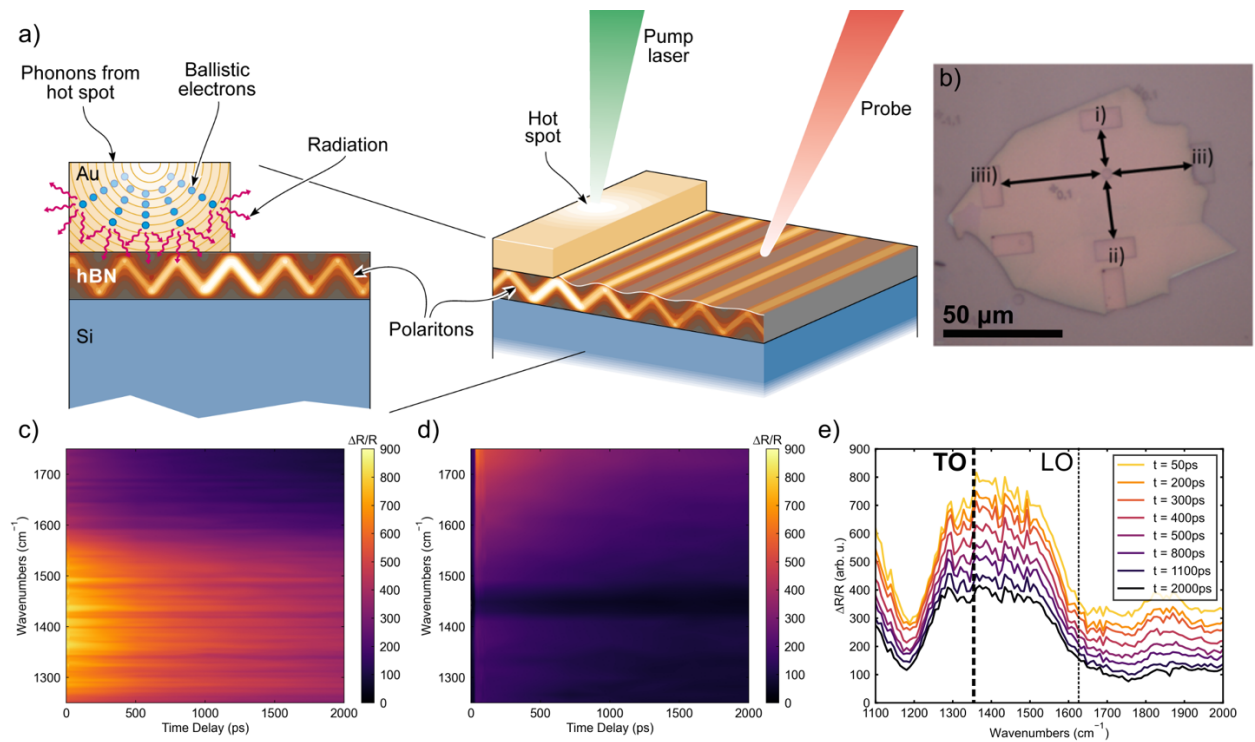

*Figure 1: Experimental details and spectral-temporal response of HPhP modes in hBN.  a) Illustration of the proposed mechanism and experimental measurement. A pump pulse (520 nm) heats a gold pad, while a sub-picosecond tunable mid-IR probe pulse measures the modulated reflectivity response of the hBN patterned flake. After pulse absorption in the Au, both phonons and ballistic electrons spread from the hot spot in the Au, depicted by the small blue particles and the background waves emanating from the hot spot. The hot carriers begin to radiate as they move and as they deposit their energy at the interface, during which radiation (the red arrows) escapes and couples into the HPhP modes of hBN. (pump and probe spot sizes not to scale) b) The sample geometry showing the position and separation of the excitation pad and the reflector pads (separation distances of i) 20 μm, ii) 30 μm, iii) 40 μm, iiii) 50 μm), all of which surpass typical HPhP propagation lengths. The excitation pad is a 50-nm-thick gold disk of radius 1.5 μm. The reflector pads are all 15 x 5 μm rectangles with a thickness of 50 nm. c) The measured thermoreflectance signal of the 116-nm hBN flake as a function of probe energy and pump-probe delay time for an incident pump fluence of 95.5 J m$^{-2}$. The strong ΔR/R response within the Reststrahlen band and near the TO phonon frequency of hBN shows the high spectral activity within the region that can be attributed to direct launching of HPhPs from near-field radiation from the heated Au-pad, thereby allowing ultrafast thermal dissipation across the Au/hBN interface. d) For reference, a similar pump fluence of an uncoated (no Au) hBN flake is provided, noting that in this case, no temporal thermoreflectance response is observed within the range of the hBN Reststrahlen band, illustrating the critical role of the Au pad as a thermal transducer in this experiment. The dark band that appears in the middle of the blank hBN contour is attributed to the C-H group vibrational mode in PMMA. A PMMA residue is expected to be on the entire*

*sample in the regions not coated with gold originating from its use an as a photo-resistive polymer in the lithographic patterning process. e) Waterfall plots of the data shown in c) at a variety of pump-probe time delays (50-2000ps) following transient Au heating, indicating more clearly the ultrafast optical response surrounding the TO phonon mode and within the hBN Reststrahlen band (indicated by the span of the dotted lines).*

The observation of a strong thermoreflectance signal within the Reststrahlen band at picosecond timescales indicates the role that HPhPs play in this thermal dissipation process. As evidence of this, when we tune our probe energy to frequencies above the Reststrahlen band of hBN following the pumping of the adjacent gold pad, we see minimal temporal changes in the probe thermoreflectance signal (Fig. 1c, e; see dashed vertical lines to designate the TO phonon mode and the surrounding active region). However, as the probe energy is tuned to energies within the Reststrahlen band, the thermoreflectance signal exhibits large increases in the thermoreflectance that is maximized at the earliest times, indicative of large optical phonon temperature changes[54] (with an additional peak at the TO phonon energy – dark vertical dashed line in Fig. 1e). This significant increase in thermoreflectance drops off at frequencies below the TO phonon, with the broadening below this band potentially due to the deeply subwavelength modes that can also be stimulated in hBN due to the exceptionally high refractive index within this spectral range, as previously discussed for Mie resonators in 4H-SiC nanopillars.[55] The hBN dielectric function does not vary with thickness in this spectral region.[56,57] To support this posit, similar observations were performed measurements on hBN flakes with varying thicknesses ranging between 89 nm and 195 nm, as presented in the Supplementary Information; as expected, no size effects are observed.

The sample was constructed such that only ~6% of the probed region is covered with gold (see Fig. 1b), effectively diminishing any effect that it might have on the measured thermoreflectance. Yet, we observe a spectrally local thermal event occurring in the Reststrahlen band of hBN. Our lock-in detection scheme isolates any optical effects of the probe. Further, the polaritonic modes in the hBN cannot be directly launched due to the lack of a momentum matching condition from the 520 nm pump. Due to the long time constant of phonon conduction (1000's of ps) compared to the decay time of our observed modal heating, we posit that the Au-to-hBN interfacial heat transfer mechanism is driven by evanescent pumping of the hyperbolic modes in the hBN flake from the radiating Au pad. This mechanism is mediated by the excited carriers in the Au pad. Due to the short timescale of our observed modal heating (10's of ps), we assert that these carriers are the ballistic electrons in the Au that release their heat radiatively at the Au-hBN interface, initiating the $\Delta T$ required for radiative transfer.

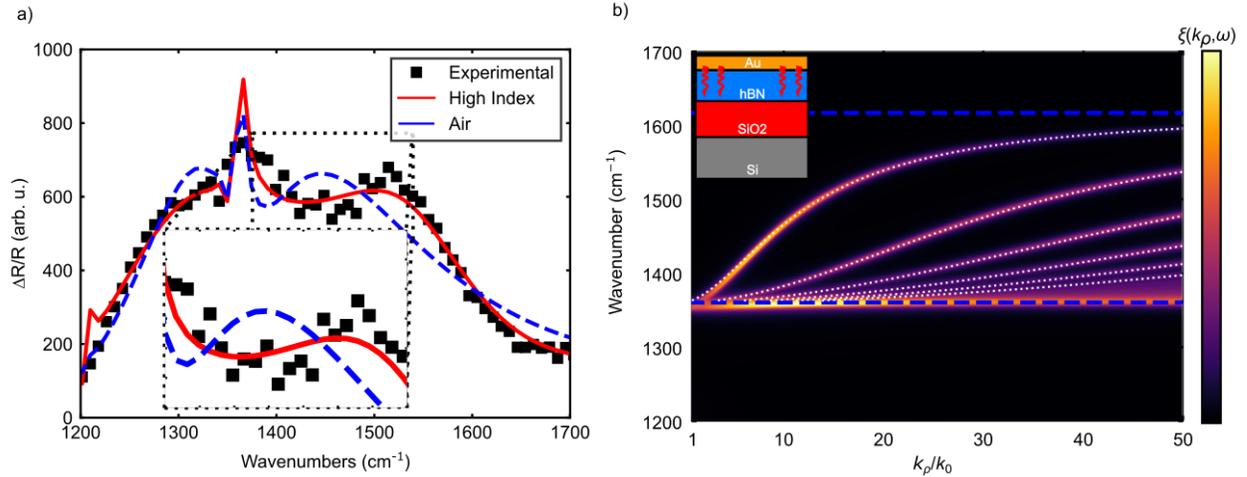

*Figure 2: Spectral response of hBN after heating from radiating Au pad. a) spectral cross sections from the contour of the 116-nm flake (Fig. 1e), fitted with and without a polaritonic coupling layer for pump-probe delay time of 10 ps. The spectral-temporal range is representative of the entirety of the activity surrounding the TO phonon mode and shows the improvement of the "Polaritonic" model compared to the standard ambient model ("Air"). The inset plot highlights the advantage of using the high-index layer for momentum matching in the TMM calculation. This "high-index layer" is a better analog to the near-field radiation from the central gold pad effectively increasing sensitivity to the TO phonon band and the polaritons therein. b) Calculated spectral heat flux per unit wavevector between Au and hBN layers in contact. The inset is a schematic of the simulated stack. The white dotted lines show the dispersion relation of HPhPs in hBN, and the blue horizontal lines show the upper and lower limits of the hBN Reststrahlen band. The fluctuational electrodynamics (FED) derived contour directly shows the majority of radiative heat flux across the interface is dominated by modes at the same momentum as HPhPs and thus satisfying our proposed launching condition. This plot begins at a normalized wave vector of 1 which means that all representative momenta are considered in the evanescent regime.*

To further explore the role of HPhPs in the ultrafast interfacial thermal transport between Au and hBN discussed above, we examine the thermoreflectance changes as a function of probe energy measured at short (10 ps, Fig. 2a) pump-probe delay times. We employ the transfer matrix method (TMM,[58] details in Supplemental Information) to simulate the thermoreflectance spectra in the vicinity of the Reststrahlen band using two models. First, we simulate this differential reflectance spectra by employing a plane wave excitation ("Air"), which lacks the high-momenta modes sufficient to launch HPhPs in hBN (Fig. 2a, dashed yellow line). In contrast, we also perform these TMM calculations using a fictional high-index prism to increase the momentum of optical modes to enable the direct launching of HPhPs (Fig. 2a, red line)[59]. This bending of the light line is effectively a simulation of the reflectivity from an attenuated total reflection (ATR) module of an FTIR system which has been experimentally shown to launch polaritons[60,61]. This simulation is therefore an approximation of the reflectivity of a system while there are active polaritons in the system. A full list of fitted parameter perturbations is available in Section B of the Supplementary Information. While no prism is employed in the experiments, this is employed as a simple method for simulating the role of the dramatically increased momenta of light, akin to using local dipole coupling to illustrate the role of polaritonic modes within this process. Note that this polaritonic-driven model provides a significant improvement in the fit as compared to the same model without the momentum-matching conditions ("Air"), providing further evidence for the role that HPhPs play in this thermal dissipation process. This effect is significantly muted at longer timescales, whereby other thermal dissipation processes such as phonon-mediated conduction across the Au/hBN interface begin to dominate (See Section B of Supplementary Information).

To determine the exact radiative conditions at the Au-hBN interface explored in Fig. 2b, we explored the distribution of the spectral radiative heat flux over different momenta. Using the

framework of FED[62], the radiative heat flux between the Au pad and hBN slab can be calculated as:

$$q''_\omega = \frac{\Theta(\omega,T_{Au})-\Theta(\omega,T_{hBN})}{4\pi^2} \Sigma_{\gamma=TE,TM} \left( \int_0^{\pi/a} \xi^\gamma(k_\rho,\omega) dk_\rho \right) \quad (1)$$

where $k_\rho$ is the parallel component of wavevector, $\Theta$ is the mean energy of an electromagnetic state, $a$ is the lattice constant of hBN, and $T_{Au}$ and $T_{hBN}$ are the temperatures of Au and hBN layers, respectively. Also, $\xi^\gamma$ is the energy transfer function, representing spectral radiative heat flux per unit $k_\rho$, between gold and hBN for $\gamma$ (TE or TM) polarization. Assuming one-dimensional (i.e., infinitely long) layered media, the energy transfer function is numerically calculated using scattering matrix method and dyadic green's functions[63]. The contour in Fig. 2b shows the energy transfer function between the gold and hBN layers. The white dotted lines in Fig. 2b also exhibits the dispersion relation of the HPhPs in hBN, found from the following equation[11]:

$$k_\rho = Re\left\{-\frac{\psi}{d}\left[\tan^{-1}\left(\frac{\varepsilon_{Au}}{\varepsilon_\perp \psi}\right) + \tan^{-1}\left(\frac{\varepsilon_{SiO_2}}{\varepsilon_\perp \psi}\right) + \pi l\right]\right\} \quad (2)$$

where,

$$\psi = \frac{\sqrt{\varepsilon_\parallel}}{i\sqrt{\varepsilon_\perp}} \quad (3)$$

Also, $\varepsilon_{Au}$ and $\varepsilon_{SiO_2}$ respectively represent the dielectric functions of gold and silica, $\varepsilon_\parallel$ and $\varepsilon_\perp$ are respectively the parallel and perpendicular dielectric functions of hBN, and $l$ is an integer number corresponding to the order of the HPhP modes. Fig. 2b shows that the radiative heat emitted by gold is predominantly received by HPhPs in hBN as the heat flux mostly occurs at wavevectors that precisely match the dispersion of HPhPs. As such, the radiative heat flux at any temperature differential between gold and hBN will be carried predominantly through HPhPs.

Beyond the spectral response, it is also critical to quantitatively analyze the temporal relaxation of the thermoreflectance signals within and outside of the Reststrahlen band to delineate the role that HPhPs play in this process. The stark contrast between the time-dependent thermoreflectance signals under these two distinct conditions are provided as a function of pump-probe delay at a probe energy approximately "on resonance" (Fig. 3a, black squares) with the TO phonon (probe energy of 7.4 μm = 0.17 eV = 1,351 cm$^{-1}$) compared to the signal with a probe energy "off resonance" (Fig. 3a, pink triangles), which is spectrally separated from the Reststrahlen band of hBN (probe energy of 6 μm = 0.21 eV = 1667 cm$^{-1}$), where hBN is nominally transparent to the probe beam (Fig. S3 in Supplementary Information). Additionally, at these photon energies, the thermoreflectance changes in Au provide nearly an order of magnitude smaller thermoreflectance signal compared to that measured from the hBN within the Reststrahlen band, thereby precluding any significant transient response from Au in driving the observed response. Thus, as the large bandgap of hBN (~5.95 eV)[64] precludes direct absorption of the incident pump energy (2.38 eV), the measured thermoreflectance signal is instead driven by the changes in the temperature of the hBN that are the result of thermal transport from the heated gold across the interface. The dominant signatures at picosecond timescales occur within the Reststrahlen band of hBN, illustrating the strong correlation of these excitations with thermal transport mediated by optical phonons and HPhPs. Further, our control measurement of pumping and probing directly on the hBN in the absence of Au results in a negligible response (Fig. S4 in Supplementary Information), clearly indicating that the measured signal (Fig. 3a) is due to a remote heating effect initiated via pump absorption within the Au pad. Thus, the correlation between the ultrafast heating of Au with the dramatic changes in the hBN thermoreflectance within the Reststrahlen band at ps timescales clearly illustrates that the mechanism is mediated through the launching of HPhPs via near-field thermal radiation from the heated gold pads. Our results therefore qualitatively suggest that the Au/hBN thermal boundary conductance (TBC) can be significantly influenced by energy transfer across the Au/hBN interface at ultrafast time scales (sub-nanosecond) mediated by HPhPs.

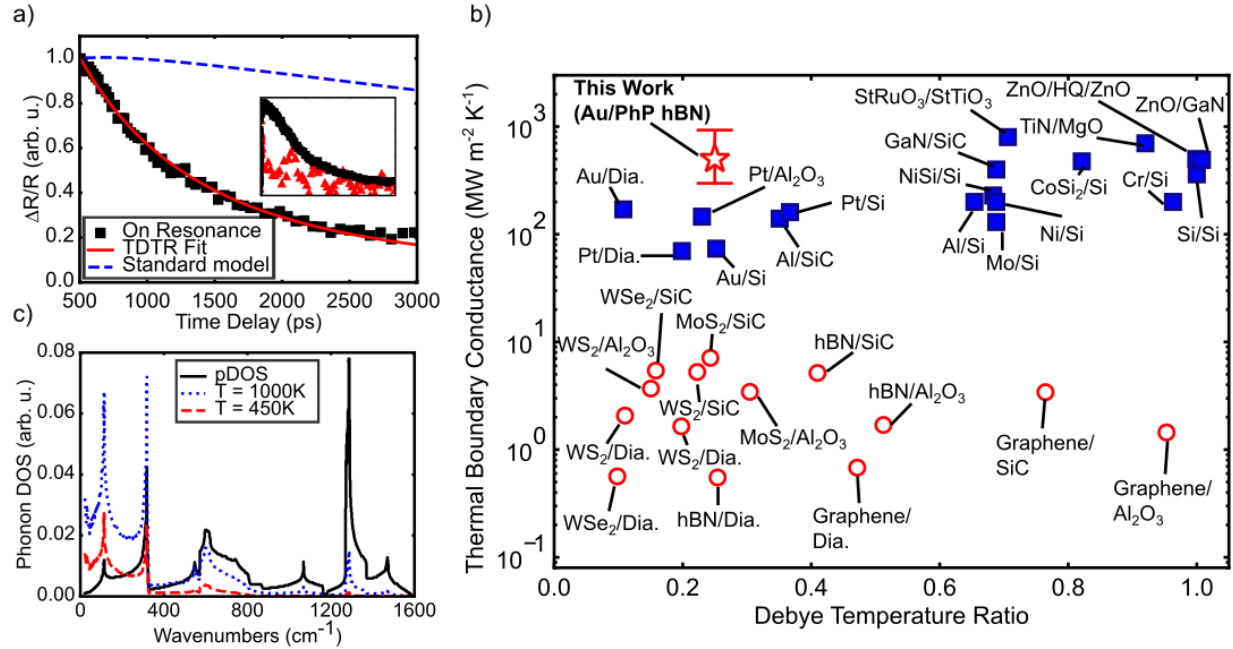

*Figure 3: Results from thermal analysis of IR Thermoreflectance* a) Thermoreflectivity response of hBN as a function of pump-probe delay time after Au pad heating near the TO resonant frequency (black squares, 7.4 μm) of hBN alongside the analytical model fit to the on-resonance data(red solid line). The best fit for the on-resonance data shown resulted in a HPhP-mediated thermal boundary conductance of >500 MW m$^{-2}$K$^{-1}$. The standard model (blue dashed line) shows the calculated thermoreflectance signal expected at the surface of the Au pads assuming literature thermal parameters as well as a Au/hBN phonon-phonon thermal boundary conductance of 12.5 MW m$^{-2}$K$^{-1}$ measured with TDTR (see Supplementary Information for details) the inset shows a comparison of the raw signal magnitude on resonance (black squares, 7.4 μm) to off resonance(red triangles, 6 μm); b) The current state of experimentally measured bulk thermal boundary conductances across 3D/3D material interfaces (filled blue squares)[2] as well as predicted 2D/3D interface conductances (open red circles region)[65] as well as the minimum Au/hBN HPhP thermal boundary conductance measured in this work, all plotted against film to substrate ratio of Debye temperatures. c) The phonon density of states for hBN was reproduced from a figure in Cuscó et al.[66] using Density Functional Theory (DFT) plotted with the occupied density of states the at two temperatures showing the lack of activity in the TO Phonon mode 150 K above the ambient temperature, implying that the measurements in this work are due to optical phonon activity measured via IR probing, and not from thermally excited phonon modes from conduction alone.

The data on resonance with the hBN Reststrahlen band shown in Fig. 3a exhibits an exponential decay with a time constant of ~1,300 ps. After laser pulse absorption of the gold and the resultant electron-phonon equilibration and thermalization, the majority of the temporal thermal decay is driven by the discharge of energy from the thermalized Au pads into near-field thermal radiation and direct launching of the HPhPs in the hBN. The thermal boundary conductance across the Au/hBN interface from this process can be approximated by $C_v d \tau^{-1} \approx 100$ MW m$^{-2}$ K$^{-1}$, where $C_v$ is the heat capacity of the gold, $d$ is the Au film thickness, and $\tau$ is the time constant of the thermal decay.[67] While this calculation is approximate, it suggests that the heat transfer mechanisms that describe the on-resonance data take place at a rate that is nearly an order of magnitude faster than the phonon-phonon driven TBC across the same Au/hBN (12 +/- 2 MW m$^{-2}$ K$^{-1}$ on the same hBN sample) measured using time-domain thermoreflectance (details in Section D of the

Supplementary Information). To quantify this Au/PhP hBN thermal boundary conductance more rigorously, we solved the analytical solution to the cylindrical heat equation commonly used to fit standard TDTR data,[68] and fit this solution to our data. This solution predicted a TBC of at least 500 MW m$^{-2}$ K$^{-1}$, restricted to a lower bound due to sensitivities (details in Section C of Supplementary Information). This Au/PhP hBN thermal boundary conductance is roughly 1-2 orders of magnitude higher than phonon-phonon TBCs measured across a plethora of 3D and 2D material interfaces (Fig. 3b).

It is of note that our model (shown in Fig. 3a) predicts a ~300x increase in effective in-plane polariton conductance in the hBN, consistent with the literature showing enhanced thermal transport in the in-plane direction of hBN due to polaritonic coupling[27–29]. These relative increases in local polaritonic conduction can be reasoned based on examining the maximum allowed near-field heat flux, $q"_{max}$, which is predicted as[69]:

$$q"_{max} = \frac{k_b^2 \left(\frac{\pi}{a}\right)^2}{48\hbar}(T^2_{emitter} - T^2_{absorber}) \tag{4}$$

where the maximum flux allowed in the near field for a Au/hBN interface can lead to a maximum TBC ($G_{max} = \frac{q"_{max}}{T_{Au}-T_{hBN}}$) of 2.5 GW m$^{-1}$ K$^{-1}$. Thus, with the coupling of emitted Au energy into hBN hyperbolic modes and non-ideal transfer within this process, the predicted TBC of 500 MW m$^{-1}$ K$^{-1}$ is reasonable. In the same regard, the group velocity of the local phonon modes which harbor HPhPs is given by the slope of the dispersion[66] which is near zero at the zone center. In contrast, the velocities of the launched hyperbolic modes can be approximated similarly from their dispersion[11]. From Kinetic Theory[70], the thermal conductivity of a system is directly proportional to these velocities, and as the group velocities of hyperbolic modes approach large percentages of the speed of light, the radiative energy flow within these modes will be transferred by the polaritons, thus increasing heat transfer. This enhanced thermal boundary conductance is also supported using a fit to the solution of the 1D heat equation described previously (details in Section C of Supplementary Information),[71] further supporting that the value for Au/hBN thermal boundary conductance is quantifiably larger than that due to phonon-phonon conduction alone, and is not a derivative of the assumptions used in our data reduction. We repeat these measurements and fits on three different Au patterns across two different hBN samples, with a consistent enhancement in Au/hBN thermal conductance observed among all "on-resonance" data.

Measurements of polaritonic launching are often performed directly with the use of a scattering-type scanning near-field microscopy (s-SNOM) which allows for the direct excitation and measurement of polaritonic modes[12,18,20-22,37,59]. However, by the nature of measurements in the near field, the system requires careful alignment and interpretation to isolate the effect of the tip launching and nearby reflecting sites. Using far-field optics, our measurement scheme is able to lock-in to the thermal event in the Au pad and probe the thermal trace of the polaritons through changes in reflection of the Brillouin zone center transverse optical modes and neighboring scattering sites (details in section A of the Supplementary Information). Thus, our measured signal is isolated temporally from any optical effect caused by the probe and, due to the high Debye temperature of hBN[72], the TO phonon would be frozen out at even the maximum lattice temperature predicted by conduction (Fig. 3b). This confluence of factors points directly to the launching of heat carrying thermally excited polaritonic modes.

Our solution to the heat equation predicts a maximum temperature rise in the hBN of approximately 150 K under these experimental conditions. While the absolute value of this increase is certainly prone to uncertainties in assumed input values, it is still orders of magnitude lower than temperatures required to thermally activate a substantial portion of the high frequency TO phonon and HPhP modes, and thus during the experimental measurements, the optical modes are nearly entirely frozen out from conductive lattice heating alone (Fig. 3c). The temperature needed to obtain a sufficient population of these modes is upwards of 1000 K, which is clearly not the case in our experimental measurements. Thus, this implies that the HPhPs are not excited simply by laser induced changes in the thermal populations, and instead are stimulated non-thermally through direct launching from near-field radiation emitted from the hot Au pad, further supporting our observations of indirect polaritonic coupling, and resultant energy transfer via HPhP-driven TBC.

The above observations coupled with the prior demonstrations of direct launching of propagating phonon polaritons via thermal emission reported by Greffet et al.[42] and later by Lu et al.[73,74] indicate instead a different model whereby HPhPs serve as the dominant thermal transport mechanism at early timescales. Specifically, we propose the following process. First, the ultrafast visible pump is absorbed by the gold pad, which increases its temperature causing it to radiate. Consistent with phonon-polariton launching via thermal radiation,[42,74] the local incoherent dipole moment of the thermal radiation provides the energy- and momentum-matching to directly launch HPhPs within the hBN flake. This in turn stimulates a broad spectrum of HPhP (and TO phonon) modes. Once launched, due to the part optical phonon, part light nature of the HPhP quasiparticles, these modes can carry the thermal energy away from the heat source due to the high heat capacity of the former and the high group velocity (with respect to acoustic phonons) of the latter. A similar mechanism has been recently demonstrated in Pan et al.[75] where under steady state conditions, the thermal conductivity is enhanced by to surface confined PhPs in silicon carbide. Our work extends these findings to volumetric hyperbolic modes and into the ultrafast regime. This ultrafast mechanism is of critical importance, as while the HPhPs will eventually decay back to conductively heat the local hBN lattice and thus, thermalization of the energy will result in uniform heating of the flake, at ultrafast timescales this provides opportunities for extracting the heat from localized hot-spots before catastrophic processes occur (e.g., device failure). Our results should lay the groundwork for a new generation of photonic sources and an efficient transport mechanism for mode specific heat sinks in high frequency electronics.

Here, we have experimentally established the potential for ultrafast, thermal transport across solid-solid interfaces via the transduction of thermal energy from a transiently heated gold pad into HPhPs supported within hBN. This mechanism provides insight into the role that polaritonic modes can offer in the realm of interfacial heat transfer, overcoming the traditional limitations of phonon-dictated TBC. Specifically, we demonstrate that polaritonic coupling can facilitate the optical modes to move heat across and away from a Au/hBN interface over an order of magnitude faster than acoustic phonon conduction in the same system. These investigations provide the initial forays into understanding the fundamental guiding principles of such ultrafast transport phenomena. Thus, through further studies of polariton-phonon scattering rates and a quantification of energy tunneling through different interlayers, we hope to elucidate the effectiveness of this energy transfer, including quantifying the maximum heat flux and capacity. The impact of this new mechanism has two main features that stand out: speed and controllability. Once the thermal energy is being carried by a polariton, it is travelling at ultrasonic speeds, which means that for high power or high frequency electronic devices that accumulate heat via joule heating, this mechanism can remove heat faster than it is accumulated. This would help next-generation devices to maintain safe operating temperatures, even at higher current densities or near local defect-induced hot-spots. The ability to design polaritonic modes through selection of materials and patterning of optical devices also allows for several advancements in novel computers, and thermal management. This new mechanism also serves to employ thermal energy to launch polaritons, meaning that photonic circuits may be able to have a useful nanoscale source for photons. Thermally, these results let us design and optimize new devices "shielded" from evanescent radiation by polaritonic absorbers.

Author contributions: The project was conceived by J.D.C., P.H. and J.P.M. and all experiments were equally advised by all three. The isotopically enriched hBN crystals were grown at Kansas State University by J.L. while under the advisement of J.E. Samples were fabricated by M.H. and J.M. at Vanderbilt University within the Vanderbilt Institute for Nanoscale Science and Engineering (VINSE). Infrared spectroscopic and nano-optic probe characterization of hBN flakes was performed by K.D.G., M.H. and J.M. Far-field tunable mid-infrared pump-probe experiments as well as optical and temporal modelling were performed at the University of Virginia ExSITE lab by W.H., D.H., J.T., S.Z., T.P., and P.H.

Acknowledgements: J.M. acknowledges funding from the National Science Foundation under grant NSF-DMR-1904793. K.D.-G., W.H., J.A.T, S.Z., T.P., and P.E.H were supported by the Army Research Office under grant number W911NF-21-1-0119, while J.D.C. and M.H. were supported by the Office of Naval Research grant #N00014-22-1-2035. D.M.H was supported by the National Science Foundation grant number 2318576.

**Methods**

The hBN flakes were prepared from $^{11}$B isotopically enriched source crystals by the standard mechanical exfoliation process onto oxygen plasma treated silicon substrates. As reported elsewhere[12,76,77], the hBN source crystals were produced by precipitation from a molten metal solution using isotopically enriched precursors (nominally 99.4 % $^{11}$B) provided by a commercial source. To fabricate the samples, the hBN flakes were initially exfoliated onto a bilayer resist transfer wafer where they were imaged optically and via AFM to down select the ideal flakes for these experiments. Transfer of the flakes onto a standard thermal $SiO_2$ film on a silicon substrate was performed via standard approaches in a home-built two-dimensional material transfer microscope. The gold pads were patterned using electron beam lithography, development, electron-beam deposition of a 50-nm thick gold film, followed by a standard liftoff procedure in n-methyl pyrrolidone (NMP). The flake sizes were characterized by atomic force microscopy and patterned using electron beam lithography, followed by electron beam deposition of a 50 nm gold film.

The infrared thermoreflectance measurements were performed using a Spectra-Physics 30W Spirit laser (369fs pulses at 1040nm with a repetition rate of 500kHz) to seed a 16W Spectra-Physics Spirit One OPA for use as a tunable (2-16um) probe pulse, as well as an 8W HIRO SHG box acting as our visible pump. These two laser lines are directed through a standard pump advancing pump-probe system (see Supplementary Information). The pump is then focused through a 125mm focal length borosilicate lens to a 30um spot size. This is then spatially and temporally overlapped with probe pulses that have been focused through a 50mm $CaF_2$ to a 100um spot. The reflected probe is then spectrally filtered to eliminate pump bleed through, and collected by a Vigo Photonics Mercury Cadmium Telluride (MCT) photodetector. The signal voltage is then passed to a Zurich Instruments UHFLI 600MHz lock-in amplifier which amplifies the voltage with via signal boxcar and demodulated by the pump modulation frequency (optically chopped at 301Hz) into X and Y signal which is then recorded alongside the amplified boxcar auxiliary voltage as a change in reflectance and raw reflectance respectively.

**References:**

1. Swartz, E. T. & Pohl, R. O. Thermal boundary resistance. *Rev. Mod. Phys.* **61**, 605–688 (1989).

2. Giri, A. & Hopkins, P. E. A Review of Experimental and Computational Advances in Thermal Boundary Conductance and Nanoscale Thermal Transport across Solid Interfaces. *Adv. Funct. Mater.* **30**, (2020).

3. Pop, E. Energy dissipation and transport in nanoscale devices. *Nano Res.* **3**, 147–169 (2010).

4. Chen, G. Size and Interface Effects on Thermal Conductivity of Superlattices and Periodic Thin-Film Structures. *J. Heat Transf.* **119**, 220–229 (1997).

5. Giri, A. & Hopkins, P. E. Achieving a better heat conductor. *Nat. Mater. 2020 195* **19**, 482–484 (2020).

6. Barker, A. S. Infrared Absorption of Localized Longitudinal-Optical Phonons. *Phys. Rev. B* **7**, 2507–2520 (1973).

7. Adachi, S. *Optical Properties of Crystalline and Amorphous Semiconductors*. *Optical Properties of Crystalline and Amorphous Semiconductors* (Springer US, 1999). doi:10.1007/978-1-4615-5241-3.

8. Poddubny, A., Iorsh, I., Belov, P. & Kivshar, Y. Hyperbolic metamaterials. *Nat. Photonics* **7**, 958–967 (2013).

9. Caldwell, J. D. *et al.* Sub-diffractional volume-confined polaritons in the natural hyperbolic material hexagonal boron nitride. *Nat. Commun.* **5**, (2014).

10. Caldwell, J. D. *et al.* Photonics with hexagonal boron nitride. *Nat. Rev. Mater.* **4**, 552–567 (2019).

11. Dai, S. *et al.* Tunable phonon polaritons in atomically thin van der Waals crystals of boron nitride. *Science* **343**, 1125–1129 (2014).

12. Giles, A. J. *et al.* Ultralow-loss polaritons in isotopically pure boron nitride. *Nat. Mater.* **17**, 134–139 (2018).

13. Cuscó, R. *et al.* Isotopic effects on phonon anharmonicity in layered van der Waals crystals: Isotopically pure hexagonal boron nitride. *Phys. Rev. B* **97**, (2018).

14. Ma, W. *et al.* In-plane anisotropic and ultra-low-loss polaritons in a natural van der Waals crystal. *Nature* **562**, 557–562 (2018).

15. Zheng, Z. *et al.* Highly Confined and Tunable Hyperbolic Phonon Polaritons in Van Der Waals Semiconducting Transition Metal Oxides. *Adv. Mater.* **30**, (2018).

16. Taboada-Gutiérrez, J. *et al.* Broad spectral tuning of ultra-low-loss polaritons in a van der Waals crystal by intercalation. *Nat. Mater.* **19**, 964–968 (2020).

17. Passler, N. C. *et al.* Hyperbolic shear polaritons in low-symmetry crystals. *Nature* **602**, 595–600 (2022).

18. Hu, G. *et al.* Real-space nanoimaging of hyperbolic shear polaritons in a monoclinic crystal. *Nat. Nanotechnol.* (2022) doi:10.1038/s41565-022-01264-4.

19. Chen, M. *et al.* Configurable phonon polaritons in twisted α-MoO3. *Nat. Mater.* **19**, 1307–1311 (2020).

20. Pavlidis, G. *et al.* Experimental confirmation of long hyperbolic polariton lifetimes in monoisotopic (10B) hexagonal boron nitride at room temperature. *APL Mater.* **9**, (2021).

21. Ni, G. *et al.* Long-Lived Phonon Polaritons in Hyperbolic Materials. *Nano Lett.* **21**, 5767–5773 (2021).

22. Li, P. *et al.* Hyperbolic phonon-polaritons in boron nitride for near-field optical imaging and focusing. *Nat. Commun.* **6**, (2015).

23. Dai, S. *et al.* Subdiffractional focusing and guiding of polaritonic rays in a natural hyperbolic material. *Nat. Commun.* **6**, (2015).

24. He, M. *et al.* Ultrahigh-Resolution, Label-Free Hyperlens Imaging in the Mid-IR. *Nano Lett.* **21**, 7921–7928 (2021).

25. Salihoglu, H. *et al.* Energy Transport by Radiation in Hyperbolic Material Comparable to Conduction. *Adv. Funct. Mater.* **30**, (2020).

26. Wu, X. & Fu, C. Near-field radiative heat transfer between uniaxial hyperbolic media: Role of volume and surface phonon polaritons. *J. Quant. Spectrosc. Radiat. Transf.* **258**, (2021).

27. Tielrooij, K. J. *et al.* Out-of-plane heat transfer in van der Waals stacks through electron-hyperbolic phonon coupling. *Nat. Nanotechnol.* **13**, 41–46 (2018).

28. Principi, A. *et al.* Super-Planckian Electron Cooling in a van der Waals Stack. *Phys. Rev. Lett.* **118**, (2017).

29. Yang, W. *et al.* A graphene Zener-Klein transistor cooled by a hyperbolic substrate. *Nat. Nanotechnol.* **13**, 47–52 (2018).

30. Wang, H. *et al.* Ultrafast relaxation dynamics of hot optical phonons in graphene. *Appl. Phys. Lett.* **96**, (2010).

31. Tokmakoff, A., Sauter, B. & Fayer, M. D. Temperature-dependent vibrational relaxation in polyatomic liquids: Picosecond infrared pump-probe experiments. *J. Chem. Phys.* **100**, 9035–9043 (1994).

32. Kapitza, P. L. Heat Transfer and Superfluidity of Helium II. *Phys. Rev.* **60**, 354–355 (1941).

33. Chen, D. Z. A., Narayanaswamy, A. & Chen, G. Surface phonon-polariton mediated thermal conductivity enhancement of amorphous thin films. *Phys. Rev. B - Condens. Matter Mater. Phys.* **72**, (2005).

34. Thompson, D., Zhu, L., Meyhofer, E. & Reddy, P. Nanoscale radiative thermal switching via multi-body effects. *Nat. Nanotechnol.* **15**, 99–104 (2020).

35. Guddala, S. *et al.* Topological phonon-polariton funneling in midinfrared metasurfaces. *Science* **374**, 225–227 (2021).

36. Ma, W. *et al.* Ghost hyperbolic surface polaritons in bulk anisotropic crystals. *Nature* **596**, 362–366 (2021).

37. Caldwell, J. D. *et al.* Low-loss, infrared and terahertz nanophotonics using surface phonon polaritons. *Nanophotonics* **4**, 44–68 (2015).

38. Foteinopoulou, S., Devarapu, G. C. R., Subramania, G. S., Krishna, S. & Wasserman, D. Phonon-polaritonics: Enabling powerful capabilities for infrared photonics. *Nanophotonics* (2019) doi:10.1515/nanoph-2019-0232.

39. Yoxall, E. *et al.* Direct observation of ultraslow hyperbolic polariton propagation with negative phase velocity. *Nat. Photonics* **9**, 674–678 (2015).

40. Caldwell, J. D., Vurgaftman, I. & Tischler, J. G. Mid-infrared nanophotonics: Probing hyperbolic polaritons. *Nat. Photonics* **9**, 638–640 (2015).

41. Tervo, E. J., Adewuyi, O. S., Hammonds, J. S. & Cola, B. A. High thermal conductivity in polaritonic SiO2 nanoparticle beds. *Mater. Horiz.* **3**, 434–441 (2016).

42. Greffet, J.-J. *et al.* Coherent emission of light by thermal sources. *Nature* **416**, 61–64 (2002).

43. Thompson, D. *et al.* Hundred-fold enhancement in far-field radiative heat transfer over the blackbody limit. *Nature* **561**, 216–221 (2018).

44. Narayanaswamy, A., Shen, S., Hu, L., Chen, X. & Chen, G. Breakdown of the Planck blackbody radiation law at nanoscale gaps. *Appl. Phys. Mater. Sci. Process.* **96**, 357–362 (2009).

45. Fiorino, A. *et al.* Giant Enhancement in Radiative Heat Transfer in Sub-30 nm Gaps of Plane Parallel Surfaces. *Nano Lett.* **18**, 3711–3715 (2018).

46. Shen, S., Narayanaswamy, A. & Chen, G. Surface phonon polaritons mediated energy transfer between nanoscale gaps. *Nano Lett.* **9**, 2909–2913 (2009).

47. Sellan, D. P. *et al.* Phonon transport across a vacuum gap. *Phys. Rev. B - Condens. Matter Mater. Phys.* **85**, (2012).

48. Song, B. *et al.* Enhancement of near-field radiative heat transfer using polar dielectric thin films. *Nat. Nanotechnol.* **10**, 253–258 (2015).

49. Warzoha, R. J. *et al.* Applications and Impacts of Nanoscale Thermal Transport in Electronics Packaging. *J. Electron. Packag. Trans. ASME* **143**, (2021).

50. Aryana, K. *et al.* Suppressed electronic contribution in thermal conductivity of Ge2Sb2Se4Te. *Nat. Commun.* **12**, (2021).

51. Aryana, K. *et al.* Interface controlled thermal resistances of ultra-thin chalcogenide-based phase change memory devices. *Nat. Commun.* **12**, (2021).

52. Fiorino, A. *et al.* Nanogap near-field thermophotovoltaics. *Nat. Nanotechnol.* **13**, 806–811 (2018).

53. Hohlfeld, J. *et al.* Electron and lattice dynamics following optical excitation of metals. *Chem. Phys.* **251**, 237–258 (2000).

54. Tomko, J. A., Kumar, S., Sundararaman, R. & Hopkins, P. E. Temperature dependent electron-phonon coupling of Au resolved via lattice dynamics measured with sub-picosecond infrared pulses. *J. Appl. Phys.* **129**, (2021).

55. Howes, A., Nolen, J. R., Caldwell, J. D. & Valentine, J. Near-Unity and Narrowband Thermal Emissivity in Balanced Dielectric Metasurfaces. *Adv. Opt. Mater.* **8**, (2020).

56. Kumar, P., Chauhan, Y. S., Agarwal, A. & Bhowmick, S. Thickness and Stacking Dependent Polarizability and Dielectric Constant of Graphene-Hexagonal Boron Nitride Composite Stacks. *J. Phys. Chem. C* **120**, 17620–17626 (2016).

57. Laturia, A., Van de Put, M. L. & Vandenberghe, W. G. Dielectric properties of hexagonal boron nitride and transition metal dichalcogenides: from monolayer to bulk. *Npj 2D Mater. Appl.* **2**, (2018).

58. Passler, N. C. & Paarmann, A. Generalized 4 × 4 matrix formalism for light propagation in anisotropic stratified media: study of surface phonon polaritons in polar dielectric heterostructures. *J. Opt. Soc. Am. B* **34**, 2128 (2017).

59. Low, T. *et al.* Polaritons in layered two-dimensional materials. *Nat. Mater.* **16**, 182–194 (2017).

60. Folland, T. G. *et al.* Probing hyperbolic polaritons using infrared attenuated total reflectance micro-spectroscopy. *MRS Commun.* **8**, 1418–1425 (2018).

61. Bludov, Y. V. & Vasilevskiy, M. I. ATR excitation of surface polaritons at the interface between a metal and a layer of nanocrystal quantum dots. Preprint at http://arxiv.org/abs/1102.2320 (2011).

62. Francoeur, M. & Pinar Mengüç, M. Role of fluctuational electrodynamics in near-field radiative heat transfer. *J. Quant. Spectrosc. Radiat. Transf.* **109**, 280–293 (2008).

63. Francoeur, M., Pinar Mengüç, M. & Vaillon, R. Solution of near-field thermal radiation in one-dimensional layered media using dyadic Green's functions and the scattering matrix method. *J. Quant. Spectrosc. Radiat. Transf.* **110**, 2002–2018 (2009).

64. Schuster, R., Habenicht, C., Ahmad, M., Knupfer, M. & Büchner, B. Direct observation of the lowest indirect exciton state in the bulk of hexagonal boron nitride. *Phys. Rev. B* **97**, (2018).

65. Foss, C. J. & Aksamija, Z. Quantifying thermal boundary conductance of 2D-3D interfaces. *2D Mater.* **6**, (2019).

66. Gil, B., Cassabois, G., Cusco, R., Fugallo, G. & Artus, L. Boron nitride for excitonics, nano photonics, and quantum technologies. *Nanophotonics* **9**, 3483–3504 (2020).

67. Hopkins, P. E., Norris, P. M. & Stevens, R. J. Influence of inelastic scattering at metal-dielectric interfaces. *J. Heat Transf.* **130**, (2008).

68. Cahill, D. G. Analysis of heat flow in layered structures for time-domain thermoreflectance. *Rev. Sci. Instrum.* **75**, 5119–5122 (2004).

69. Basu, S. & Zhang, Z. M. Maximum energy transfer in near-field thermal radiation at nanometer distances. *J. Appl. Phys.* **105**, 093535 (2009).

70. Kittel, C. *Introduction to Solid State Physics*. (Jhon Wiley & Sons, 1996).

71. Jiang, P., Qian, X., Yang, R. & Lindsay, L. Anisotropic thermal transport in bulk hexagonal boron nitride. *Phys. Rev. Mater.* **2**, 064005 (2018).

72. Tohei, T., Kuwabara, A., Oba, F. & Tanaka, I. Debye temperature and stiffness of carbon and boron nitride polymorphs from first principles calculations. *Phys. Rev. B* **73**, 064304 (2006).

73. Lu, G., Tadjer, M., Caldwell, J. D. & Folland, T. G. Multi-frequency coherent emission from superstructure thermal emitters. *Appl. Phys. Lett.* **118**, (2021).

74. Lu, G. *et al.* Engineering the spectral and spatial dispersion of thermal emission via polariton-phonon strong coupling. *Nano Lett.* **21**, 1831–1838 (2021).

75. Pan, Z. *et al.* Remarkable heat conduction mediated by non-equilibrium phonon polaritons. *Nature* **623**, 307–312 (2023).

76. Vuong, T. Q. P. *et al.* Isotope engineering of van der Waals interactions in hexagonal boron nitride. *Nat. Mater. 2017 172* **17**, 152–158 (2017).

77. Liu, S. *et al.* Single Crystal Growth of Millimeter-Sized Monoisotopic Hexagonal Boron Nitride. *Chem. Mater.* **30**, 6222–6225 (2018).

# Supplementary Information

## Ultrafast evanescent heat transfer across solid state interfaces via phonon-polaritons

William Hutchins,[1] John A. Tomko,[1] Dan M. Hirt[1], Saman Zare[1], Joseph R. Matson,[2] Katja Diaz-Granados,[2] Mingze He,[3] Jiahan Li[4], Thomas Pfeifer[1], James Edgar,[4] Jon-Paul Maria,[5] Joshua D. Caldwell,[2,3,*] Patrick E. Hopkins[1,6,7,*]

1. Department of Mechanical and Aerospace Engineering, University of Virginia, Charlottesville, VA 22904, USA

2. Interdisciplinary Materials Science Program, Vanderbilt University, Nashville, TN 37212, USA

3. Department of Mechanical Engineering, Vanderbilt University, Nashville, TN 37212, USA

4. Tim Taylor Dept. of Chemical Engineering, Kansas State University, Manhattan, KS, USA.

5. Department of Materials Science and Engineering, Pennsylvania State University, University Park, Pennsylvania 16802, USA

6. Department of Materials Science and Engineering, University of Virginia, Charlottesville, VA 22904, USA

7. Department of Physics, University of Virginia, Charlottesville, VA 22904, USA

| Correspondence Addresses |
|---|
| Professor Joshua D. Caldwell |
| Interdisciplinary Materials Science Program |
| Department of Mechanical Engineering |
| Vanderbilt University |
| Nashville, TN 31212 , USA |
| josh.caldwell@vanderbilt.edu |
| |
| Professor Patrick E. Hopkins |
| Department of Mechanical and Aerospace Engineering |
| Department of Materials Science and Engineering, |
| Department of Physics, |
| University of Virginia |
| Charlottesville, VA 22904, USA |
| phopkins@virginia.edu |

# A. Visible Pump Infrared probe measurements

In order to study the ultrafast heat transfer processes of the optical phonons in hBN after thermal coupling with the Au pad via evanescent heat transfer, we utilize a home-built ultrafast pump-probe system with tunable probe wavelengths. Our experiment is built around a regenerative amplified laser system (Spectra Physics SPIRIT) emanating ~400 fs, 1040 nm pulses at a repetition rate of 500 kHz. The output of this laser is directed into two paths, the "pump" and "probe". The pump path is frequency doubled to 520 nm then passed down a mechanical translation stage, which changes the path length with respect to the probe pulse; in our configuration, we shorten the delay line of the pump with respect to the probe to generate our transient reflectivity data. The pump pulse train is then chopped at a rate of 451 Hz and focused onto the sample surface at the patterned flake of interest. The probe pulse gets fed through a mid-Infrared (MIR) optical parametric amplifier (OPA) which then outputs a range of frequencies from 2-16 microns. The OPA splits the seed laser into two paths the first is used as an high energy pump (520nm, the second harmonic of the seed) which is primarily used for power amplification of the seed, after this amplification step the seed is referred to as the signal beam (680-840nm) and is used in mixing to generate the desired output, the second path is sent though a white light generation (600 -1100nm, WLG) crystal and is subsequently used for the process of seed amplification. The amplification step results in additional photons which constitute the Idler beam (1350-2060nm), these three beams have an simple relationship (below) between their wavelengths given by the conservation of energy.

$$\frac{1}{\lambda_{idler}} = \frac{1}{\lambda_{pump}} - \frac{1}{\lambda_{signal}} \tag{1}$$

A similar process of amplification is repeated using the Idler (1350-2060nm) from the first amplification step, the Idler from the second amplification step and the fundamental seed (1040nm residual after the SHG), resulting in tunable Signal (1350-2060nm) and Idler (2060-4500nm) beams. This tunability is derived from the spectral overlap of the pump and WLG pulses in the first amplification step. The two beams resultant from the final amplification step can finally be combined into a difference frequency generation (DFG) crystal for tunable output from 4-16 µm. For our work, we filter the Idler and Signal out of the path using an Germanium window and use wavelength from the DFG as the probe, allowing us to access probe wavelengths at 15 nm increments through a range of 5-9 µm to measure across the Reststrahlen band in hBN. This probe is then focused onto the sample surface, centered around the pump (Figure S1).

The patterned hBN flake is oriented on a sample holder such that the probe light incident on the surface is p-polarized. P-polarized light was chosen somewhat arbitrarily to due to physical limitations on the measurement table. Transfer matrix method calculations (detailed in section B) were used to assure that the signals would be polarization independent confirmed in figures S2 and S3. The reflection of the TO observed is most similar across polarizations at 45 degrees (figure S3) so to get consistent signal an angle of 45 degrees was used in our measurements. The reflected probe signal is then measured by a MCT photo detector sensitive to our frequency range (Vigo PC-4TE-14-1x1). We lock in to the probe pulse signal at the chopper frequency of the pump as a function of pump-probe delay time. We monitor this transient thermoreflectivity at different probe wavelengths, generating the wavelength dependent transient pump-probe contour plots, shown in figures S4 and S5. Each experiment was repeated 5 times over 2 flakes with different patterning. The additional spectral sweep on an 89nm sample flake is shown in figure S6. Showing an similar ultrafast heating of the Reststrahlen band.

To better understand our measured signal several backgrounds were taken to filter out the substrate and transducer effects. These scans, summarized in figures S4 and S5, show that there is only an very minor spectrally independent response from Au at the wavelengths of interest which is consistent with the low thermoreflectance at IR wavelengths reported in literature [4]. There does seem to be some response from the Silicon substrate, but the region that this appears is near the edge of the hBN Reststrahlen band so our mechanism should be shielded from the noise caused by this. This is further supported from the data shown when the probe wavelength lies outside of the Reststrahlen band, which is indicative of the response of the silicon substrate from the thermal discharge of the Au; as expected, at the early timescales, the heating of the Si from the Au should be negligible.

## B. Optical modelling

To determine the spectral response of the hBN, Transfer Matrix method (TMM)[5] calculations were performed on simulated stacks of hBN(116nm)/ SiO$_2$(280nm)/ Si(0.5mm) with and without a prism layer assuming incident photons on the hBN. The Au pattern was neglected from the model as the addition does not add accuracy since the spectral variation of the dielectric function of Au at these wavelengths is negligible[6]. Due to the nature of our experiment, some minor changes were made to the method for a direct comparison to our measured signal.

The first change was made to mimic the thermoreflectivity response that we measure, which is a differential signal that measures the change in probe reflectivity due to a thermal perturbation. For this, we perform two separate calculations, the first being made to an unpumped stack where all optical constants are drawn from literature (see Table S1),[1] and the second calculation made to a perturbed case, where the optical parameters in the dielectric function of hBN were adjusted slightly from their "unpumped" values. For the dielectric function of hBN, a Lorentz oscillator model was used with a "TOLO" form as below[1],

$$\varepsilon(\omega) = \varepsilon_\infty \left( \frac{\omega_{LO}^2 - \omega^2 - i\Gamma\omega}{\omega_{TO}^2 - \omega^2 - i\Gamma\omega} \right) \qquad (2)$$

where ω is the frequency and the subscripts LO and TO denote the longitudinal and transverse phonon frequencies, respectively, $\Gamma$ is the damping, and $\varepsilon_\infty$ is the high frequency permittivity. The difference between our unpumped and pumped TMM calculations yield "ΔR", which is then divided by the unperturbed calculation to mimic the reported "ΔR/R" of our measurements. In order to fit these TMM predictions to our measured data, a nonlinear least squares fit was performed by tuning the optical constants of the perturbed calculation.

The second change made to a traditional TMM approach was to accommodate the anisotropic nature of hBN. The dielectric function was separated even further into $\varepsilon_\perp$, and $\varepsilon_\parallel$ with similar TOLO forms as above. These two different dielectric functions allow for a unique treatment for each of the ordinary and extraordinary axes. The calculations used followed closely the 4x4 Transfer matrix formalism outlined by Hao et al.[5]

We noted that the TMM calculations described above could not effectively represent the data at early times with any reasonable changes (S7-S15). This led to the addition of a prism layer to our stack to better encapsulate the polariton being active in the system. The prism slightly bends the light line to better momentum match to the polariton dispersion in the hBN. In the simulation this effect is incorporated with a layer of material above the hBN flake with a flat refractive index of 1.7 across the spectra. This value was settled on as it bent the incoming light enough to mesh well with our measured spectra without distorting the spectral shape too much. The result of fitting both simulations to the data at several different temporal cross sections is shown by figures S7-S15. The most significant changes required to fit the signal were in the epilon infinity of hBN as well as the TO phonon frequencies, which both are direct indications of thermal changes in the hBN.

## C. Temporal Modelling

The first model we used to fit our data was a solution to the cylindrical heat equation using code typically used for standard TDTR data[7]. Figure S16 was created by exploring the entire space of enhanced TBCs as well as effective in-plane polariton conductance that could possibly represent the data. The curves are isolines of constant mean squared error. This illustrates that the best fits lie within the solid 0.05 isoline and highlights this models' insensitivity to the thermal boundary conductance as the value increases beyond 500 as all of these values fit within 5% error. Thus this model serves as an proof for the enhancement of TBC and thermal energy transfer between Au and hBN. We have provided additional calculations in Fig. S18 for the surface temperatures of the Au and hBN in our analytical models. Without the high boundary conductance the thermal model cannot predict the extreme curvature at early times (500-1000ps). The in-plane polariton conductance is altered to match the slope trends at longer times (<1000ps). Our analysis estimates the temperature of the surface of hBN based on the solution to the cylindrical heat equation applied to a multilayer geometry, a typical approach used for analyzing pump-probe thermoreflectance experiments in the time. We modified the model to account for our experimental conditions. Laser repetition rate was adjusted from 80MHz to 500-kHz matching our seed laser repetition rate. Our system was also modulated at much lower frequencies from TDTR (300Hz). To simulate the small pumping region in our model (the gold radiator pad), we restricted the heat flow to solely cross-plane transport regulating the flux by perturbing the TBC alone. To capture the anisotropic heat transfer of the hyperbolic modes, we perturbed the in-plane thermal conductance of the hBN layer. When fitting our data, we held the cross-plane thermal conductivity constant and varied the in-plane conduction as well as the Au/hBN interfacial conductance. A least squares fitting algorithm was then performed perturbing the fit variables. In our simulations, the temperature curve associated with the hBN was spatially averaged over the size of the entire flake.

The authors recognize that this purely diffusive model is insufficient in quantifying energy transfer rates that occur on photonic timescales.

To remedy this shortcoming, we devise an two temperature model that can isolate the TBC and get us to an better approximation. In our pump-probe measurement, electrons are excited within gold with a sub-picosecond pulse and upon excitation, heat is transferred ballistically to the gold-boron nitride interface. Because electron-phonon nonequilibrium in gold has a duration on the order of tens of picoseconds and the lifetime of optical phonon modes are on the order of hundreds of picoseconds, we consider our model as a lumped capacitance model with a single temperature in both gold and boron nitride. [8]Thus, for this model we use a parabolic one step model (POS) to model the thermoreflectance data for probe energies within the reststrahlen band of hBN that is shown by the following differential equations:

$$C_{AU}(T_{AU})\frac{\partial T_{AU}}{\partial t} = -\frac{h}{d_{au}}(T_{AU} - T_{HBN}) \tag{3}$$

$$C_{HBN}(T_{HBN})\frac{\partial T_{HBN}}{\partial t} = \frac{h}{d_{au}}(T_{AU} - T_{HBN}) + \nabla \cdot (\kappa_{HBN}\nabla T_{HBN}) \tag{4}$$

$$C_{HBN}(T_{HBN})\frac{\partial T_{HBN}}{\partial t} = \nabla \cdot (\kappa_{HBN}\nabla T_{HBN}) \tag{5}$$

Where h represent the thermal boundary conductance (TBC) between phonons in gold and optical phonons in boron nitride, $d_{AU}$ represents the thickness of gold, $\kappa_{HBN}$ represents the thermal conductivity of boron nitride, and $T_{AU}$ and $T_{HBN}$ represent the temperatures of gold and boron nitride respectively.[9,10] Equation 3 describes the thermal transport in Au where we assume that the gold depth has thermalized within the pulse duration, as the ballistic penetration depth of gold is larger than the 50 nm Au used in our experiment. [11] Thus, the only term describing thermal transport in Au is the thermal boundary conductance from gold to optical phonons in boron nitride. Equation 3 describes the Au/hBN interface node. Equation 4 describes the thermal transport within hBN where we discretize the thickness of each hBN flake into .5 nm step sizes.

In order to remove source term uncertainties, we normalize the temperatures in Eqs. (3), (4), and (5), to take the following form [12]:

$$\varphi_{f,s} = \frac{T_{f,s} - T_0}{T_f(0) - T_0} \tag{5}$$

Where $T_0$ is the temperature of the film and the substrate before excitation, $T_f(0)$ is the temperature of the film immediately after excitation, and $T_{f,s}$ is the unnormalized temperature distribution. This results in the POS model taking the following form:

$$C_{AU}(\varphi_{AU})\frac{\partial \varphi_{AU}}{\partial t} = -\frac{h}{d_{au}}(\varphi_{AU} - \varphi_{HBN}) \tag{6}$$

$$C_{HBN}(\varphi_{HBN})\frac{\partial \varphi_{HBN}}{\partial t} = \frac{h}{d_{au}}(\varphi_{AU} - \varphi_{HBN}) + \nabla \cdot (\kappa_{HBN}\nabla\varphi_{HBN}) \tag{7}$$

$$C_{HBN}(\varphi_{HBN})\frac{\partial \varphi_{HBN}}{\partial t} = \nabla \cdot (\kappa_{HBN}\nabla\varphi_{HBN}) \tag{8}$$

Subject to the following conditions in replacement of a source term:

$$\varphi_{AU}(0, t_0) = 1 \tag{9}$$

$$\varphi_{HBN}(\infty) = 0 \tag{10}$$

Where $t_0$ represents the time the sample starts heating. In our model we set $t_0$ to the time of the max thermoreflectance signal for each dataset. The constants for our two-layer model can be found in Table S3 and sensitivities for each fitting parameter can be found in Figure S17.

The temporal dependence in interface-mediated radiative polaritonic heating is further solidified by analyzing the temporal changes in the experimental data within this parabolic one step model (POS) framework. For this model we needed to introduce new values to our calculations that correspond to the heat capacities of the optical phonon branches as well as thermal conductivities for each; all values used in this calculation are detailed in Table S4. A Debye model was used to approximate the boron nitride optical phonon heat capacity. This made it possible for us to only use one unknown parameter for fitting (Figure S17), an effective thermal boundary conductance (TBC) from the bulk heat carriers of gold to the optical phonon-polariton modes of hBN. We also allowed for a 20 percent perturbation in thermal conductivity when fitting in order to better capture our data.

We begin our fits at a time of t=80 picoseconds as to not be sensitive to uncertainties associated with initial laser source heating and to be sufficiently after the equilibration time of electrons and phonons in gold.[13,14] Using an estimated PBC value of 1 GW/m$^2$K in addition to our calculated optical phonon heat capacity and measured hBN flake thicknesses.[12]

Additionally, the POS predicts a temperature rise on the order of 8K. This temperature is indicative of the maximum temperature rise of the lattice in our experiment, however at these lattice temperatures the optical modes are all frozen out (shown in main text). In order to obtain a sufficient activity of the TO phonon we needed to artificially heat the system upwards of 1000 K which implies that there is some non-thermal excitation of the optical modes, which in our case is attributed to polaritonic coupling.

## D. TDTR Measurement of Au/hBN Boundary Conductance

To prepare samples for TDTR we first deposit an 50nm thick Au transducing layer via electron beam evaporation at 7 µTorr on each of 2 samples with at least one exfoliated (see methods) flake for measurement. This sample stack was chosen to isolate the Au/hBN TBC and remaining as close to the active experiment as possible. This technique, utilizes sub-picosecond pulses emanating from an Ti-Sapphire oscillator at 80MHz repetition rate. The pulses are separated in an pump path and an time delayed probe path that is reflected from the Au transducer both paths are tinted before the sample such that the pump can be filtered out before collection. The probe beam effectively measure the change in thermoreflectivity of the surface of the transducer tracking the decay of thermal energy deposited by the pump pulse. The pump line is modulated by an Conoptics electro-optic modulator in order to demodulate the ratio of in-phase and out-of-phase reflected probe signal ($-V_{in}/V_{out}$) via lock in detection at the pump modulation frequency. The signal is recorded for atleast 5ns after the pump heating event, this characterizes the thermal decay sufficiently for thermal analysis. Three modulation frequencies (1MHz, 5MHz, 8MHz) were chosen to maximize sensitivity to the Au/hBN TBC (G1) (see figure S19). S18 was calculated by solving the cylindrical heat equation varying modulation frequency and simulating the distributions of thermal gradients in the sample stack. Then by varying each of the desired fitting parameters by 0.1% and subtracting the original decay curve then normalizing by the magnitude of the perturbation allows for a direct comparison of the relative "strength" of each parameter.

Visualizing this "strength", commonly called sensitivity, allows the experimenter to select the measurement scheme to isolate desired parameters. For example, in S18 it is shown that the buried interface resistance (G2) is relatively insensitive in our sample stack thus, no matter the quality of the hBN/SiO2 interface, other parameters like G1 and the cross plane thermal conductivity of hBN (k2) can still be fit. Using two different samples of different hBN thicknesses (89nm, and 195nm) allowed us to normalize the effect of the competing resistances out and hold TBC constant across samples.

On all TDTR samples the most sensitive parameters are the thermal conductivity and thickness of the transducer, so these two values are measured rigorously after deposition via characterization of a witness sample present in the deposition. The thickness is measured via profilometry, and the thermal conductivity is measured via four point resistivity[15] and conversion to thermal conductivity using the Wiedemann Franz law[16]. The pump and probe beams are focused on the Au Transducer at $e^{-2}$ radii values of 12 and 9 μm for the pump and probe spots respectively. Each sample was measured 3 times at each modulation frequency to reduce systematic error. Then using a numerical solution to the cylindrical Heat Equation[17] a contour is fit to all nine datasets at once is found (89nm sample shown in figure S20) with fitted and assumed Thermal parameters with significant sensitivity are listed in table S4. In an effort to quantify the error of each measurement the fitting procedure is repeated after varying all sensitive thermal parameters by up to 10% and calculating the square root of the sum of all the squares of resultant errors (see table S4).

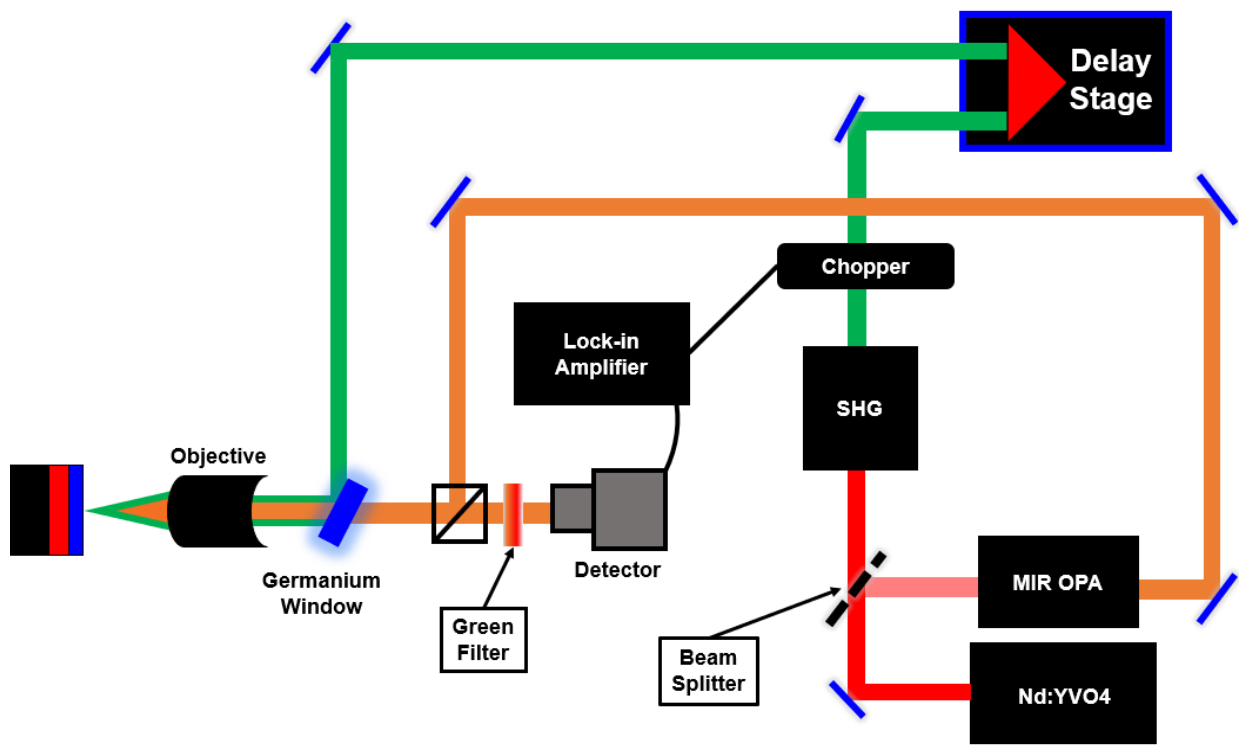

*Figure S1: A diagram of the MIR pump probe technique.*

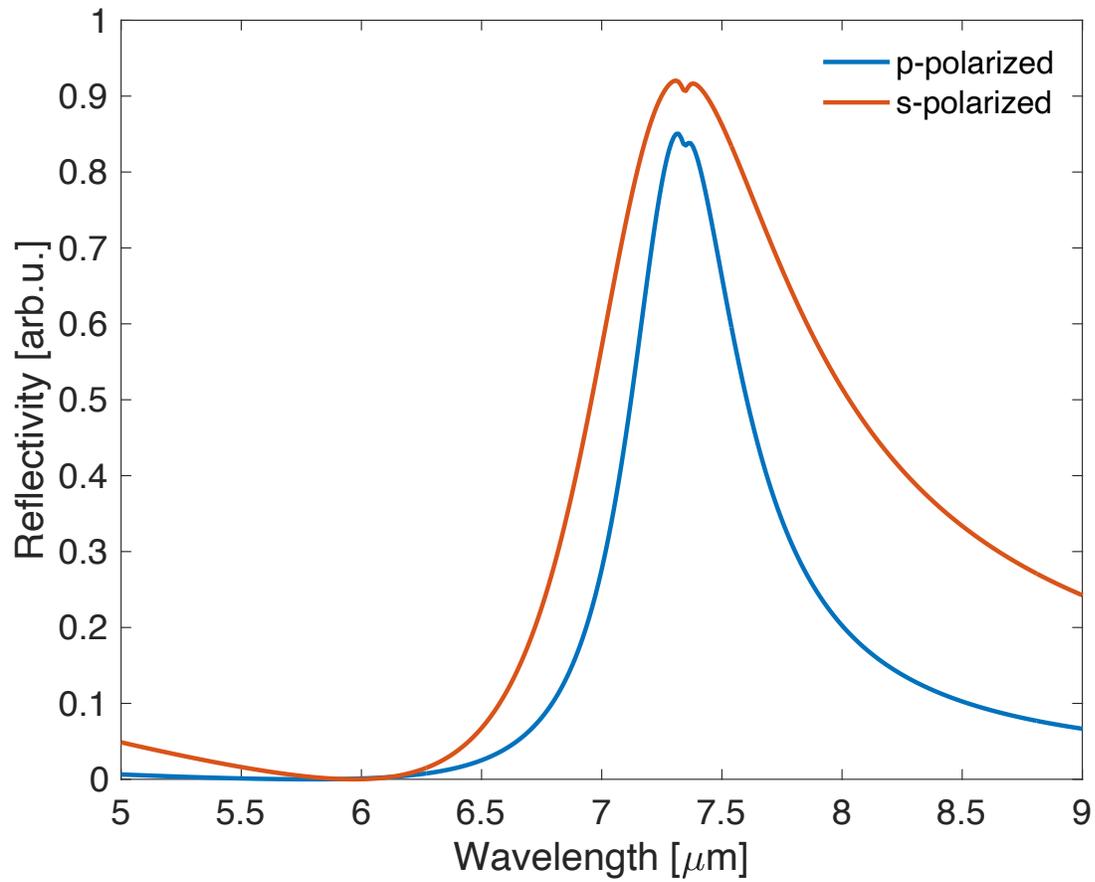

*Figure S2: TMM simulation comparing the 45 degreee reflectance of hBN within the Reststrahlen region displaying the similarity in signal.*

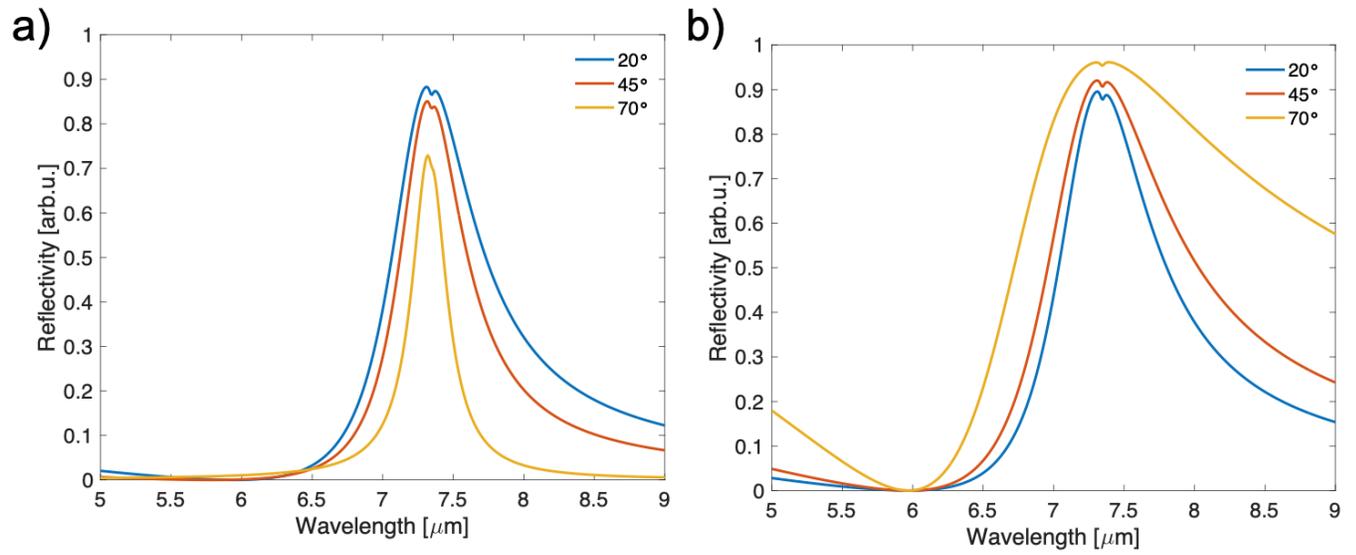

*Figure S3: TMM simulations comparing the angle depencdences for (a) p-polarization and (b) s-polarization showing commonality of signals at near 45 degree angles but high variability at angles >70 degrees.*

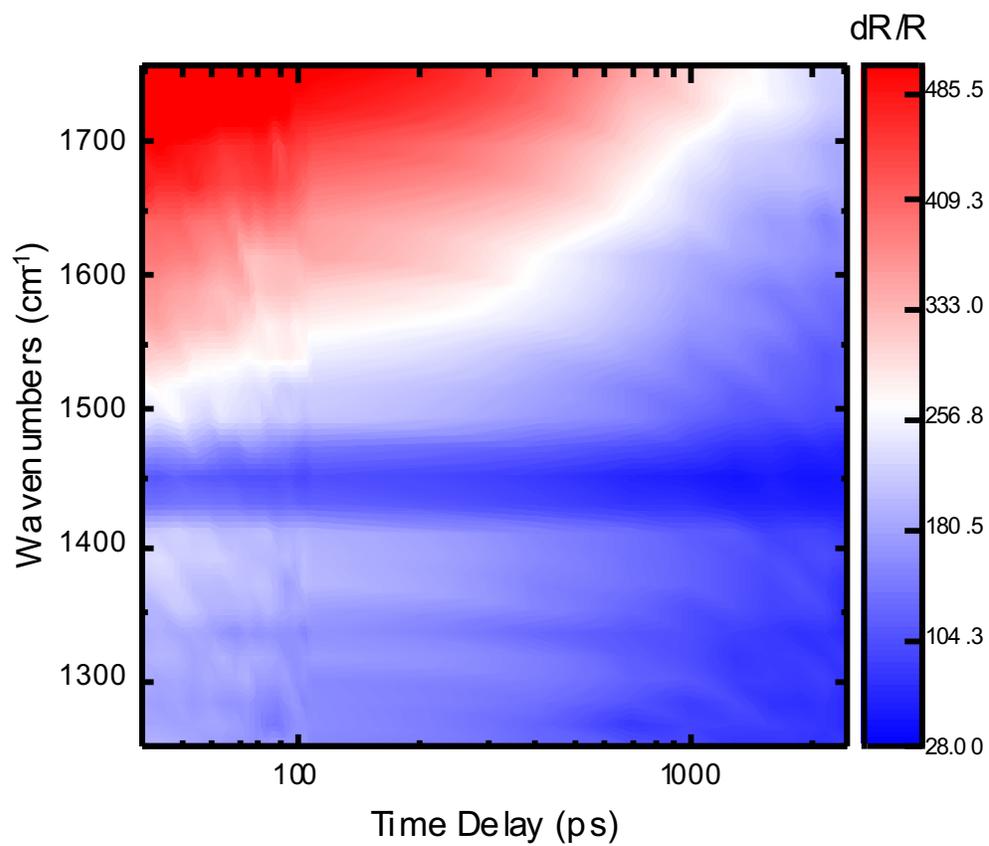

*Figure S4: A contour of a uncoated hBN flake.*

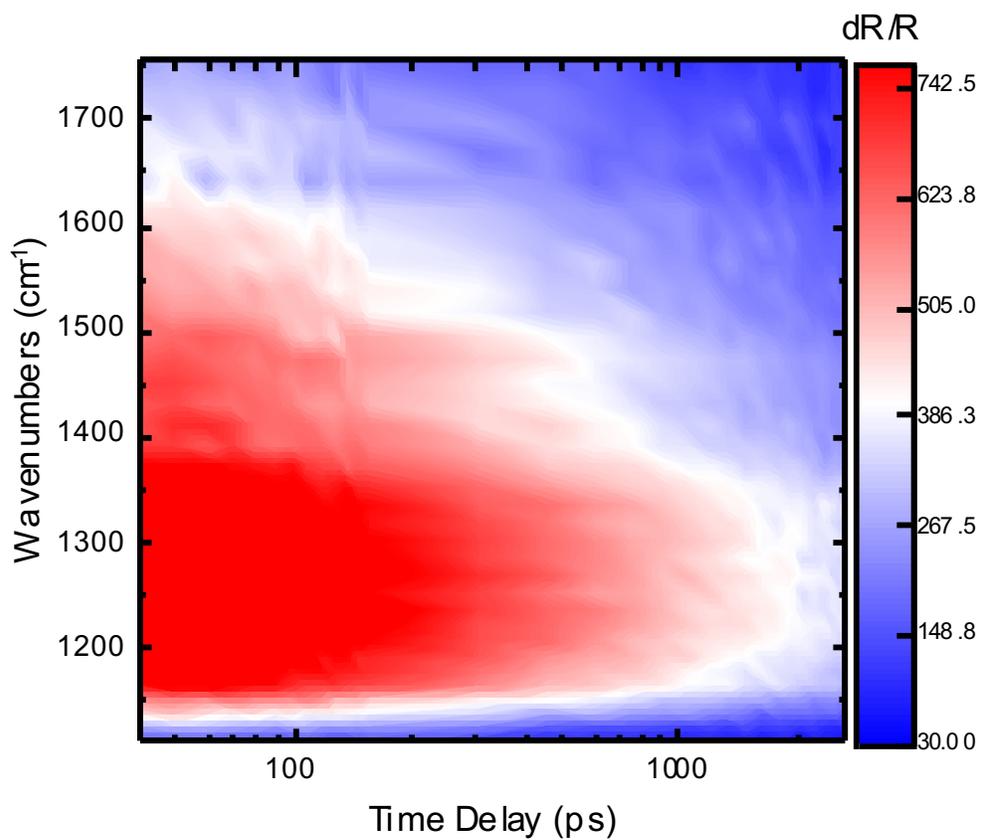

*Figure S5: This shows a sweep on the substrate beside a patterned hBN flake to determine a direct background.*

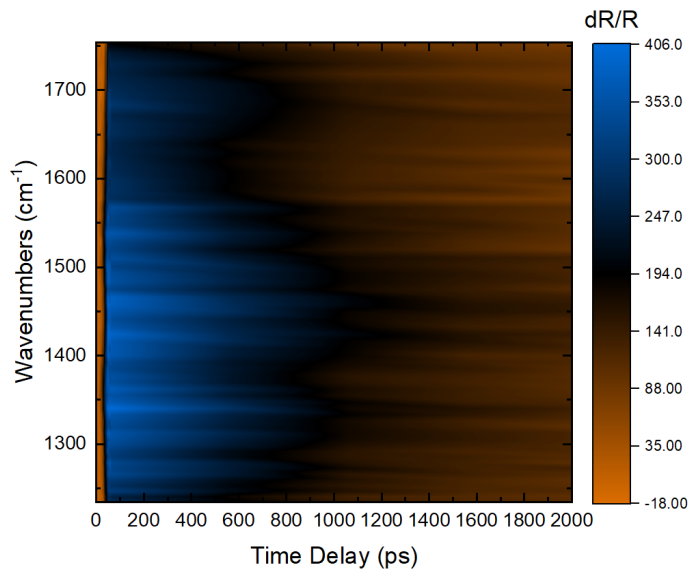

*Figure S6: Measured ΔR/R spectra taken from a separate flake with a thickness of 89 nm.*

*Table S1: Summary of unperturbed values considered in the TMM calculations. These values were also used as the starting point for the fitting procedure.*

| [11]B 99.2% | $\varepsilon_\infty$ | $\omega_{TO}(cm^{-1})$ | $\omega_{LO}(cm^{-1})$ | $\Gamma(cm^{-1})$ |
|---|---|---|---|---|
| Ordinary axis[1] | 5.32 | 1359.8 | 1608.7 | 2.1 |
| Extraordinary Axis[1] | 3.15 | 755 | 814 | 1 |

Table S2: an overview of the perturbations fitted parameters given as a percentage for each of the terms in the TO-LO form of the dielectric function for both the in plane and cross plane directions

|  | Prism(%) | Prisimless(%) |
|---|---|---|
| $\Gamma_\perp$ | 10-15 | 5-12 |
| $\Gamma_\|$ | 2-4 | 1-2 |
| $\omega_{LO\perp}$ | 0-1 | 0-1 |
| $\omega_{TO\perp}$ | 10-13 | 0-1 |
| $\omega_{LO\|}$ | 20-30 | 5-13 |
| $\omega_{TO\|}$ | 4-5 | 4-5 |

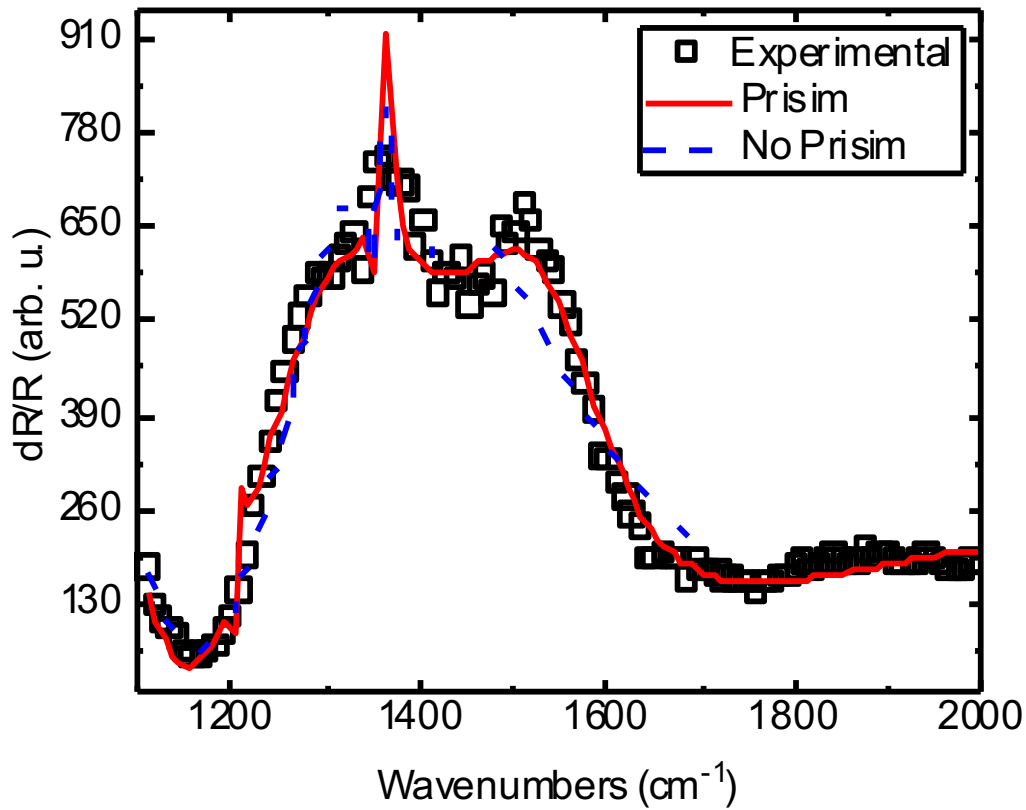

*Figure S7: The two fits with and without a prism on top of the measured signal, this figure is of a 15ps spectral cross-section.*

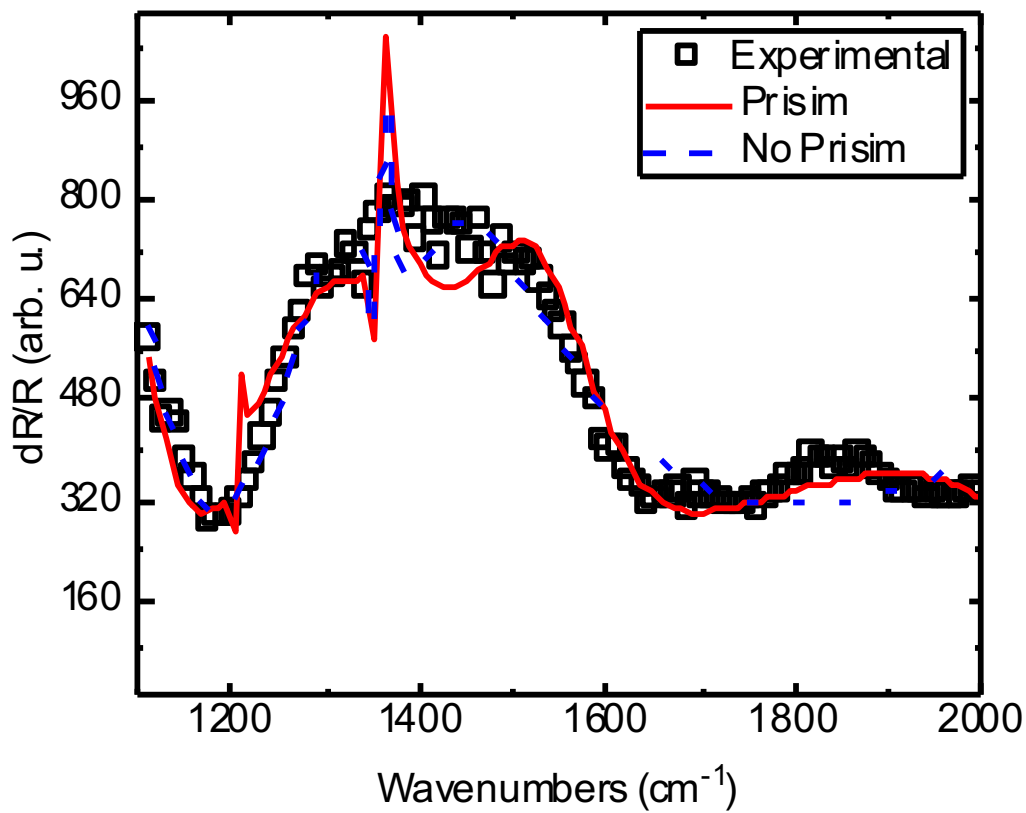

*Figure S8: The two fits with and without a prism on top of the measured signal, this figure is of a 100ps spectral cross-section.*

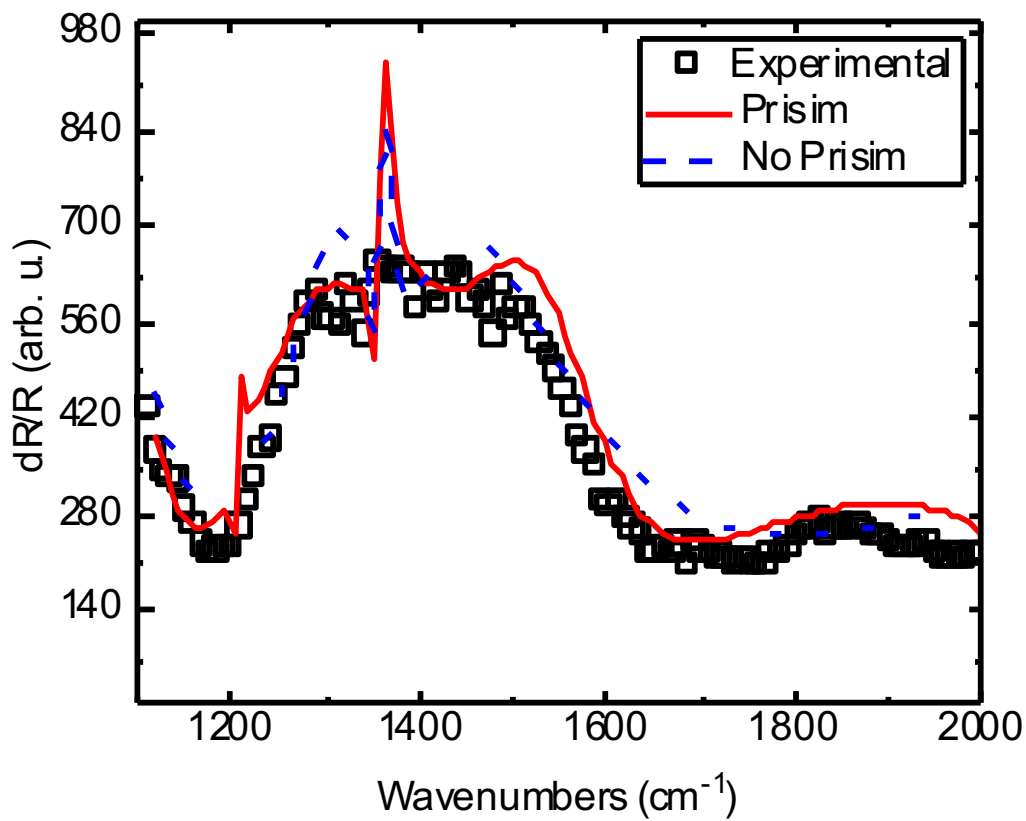

*Figure S9: The two fits with and without a prism on top of the measured signal, this figure is of a 200ps spectral cross-section.*

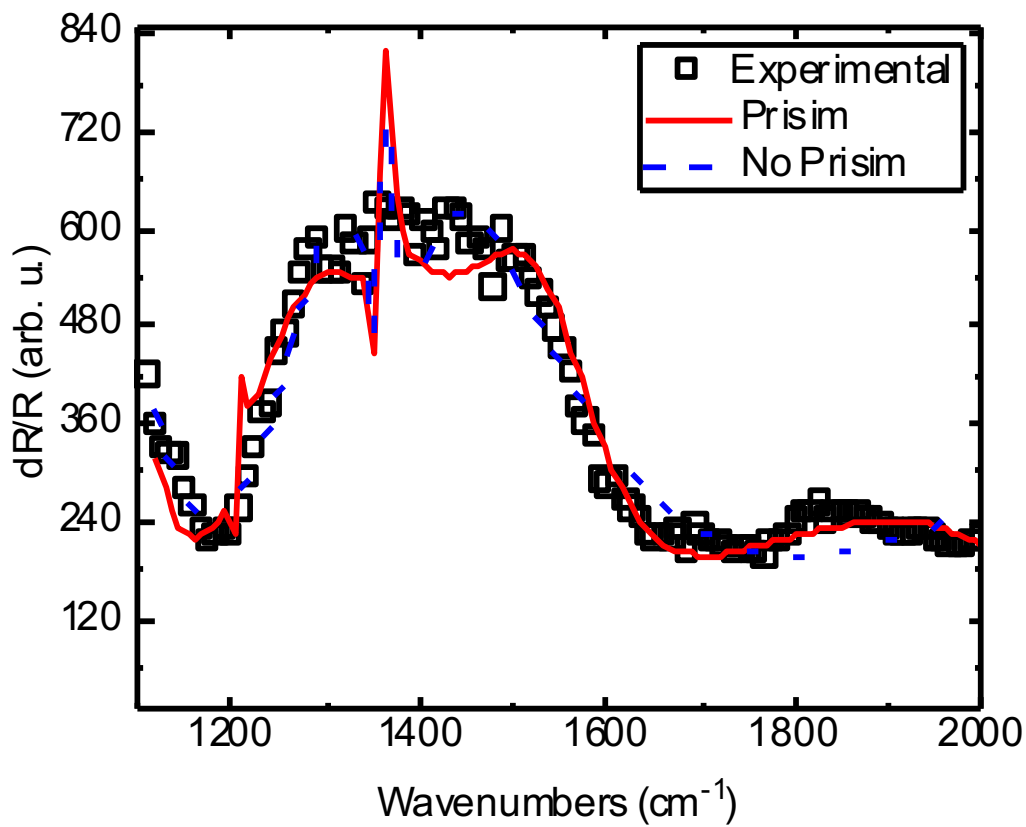

*Figure S10: The two fits with and without a prism on top of the measured signal, this figure is of a 300ps spectral cross-section.*

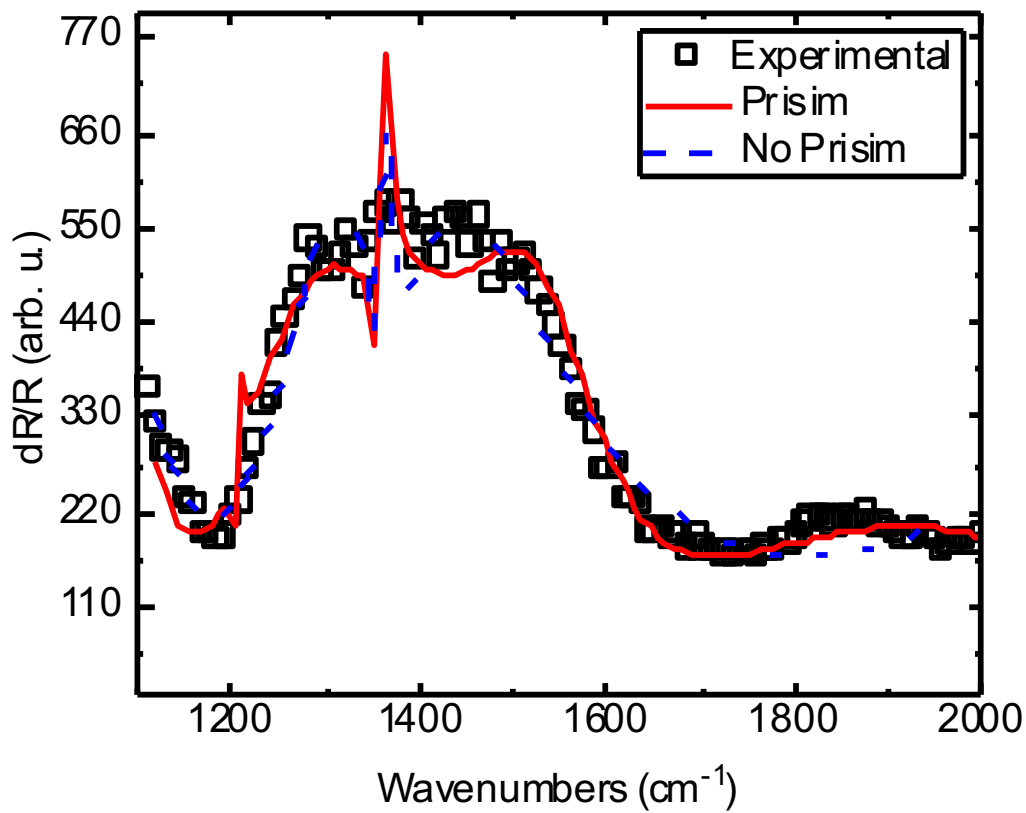

*Figure S11: The two fits with and without a prism on top of the measured signal, this figure is of a 400ps spectral cross-section.*

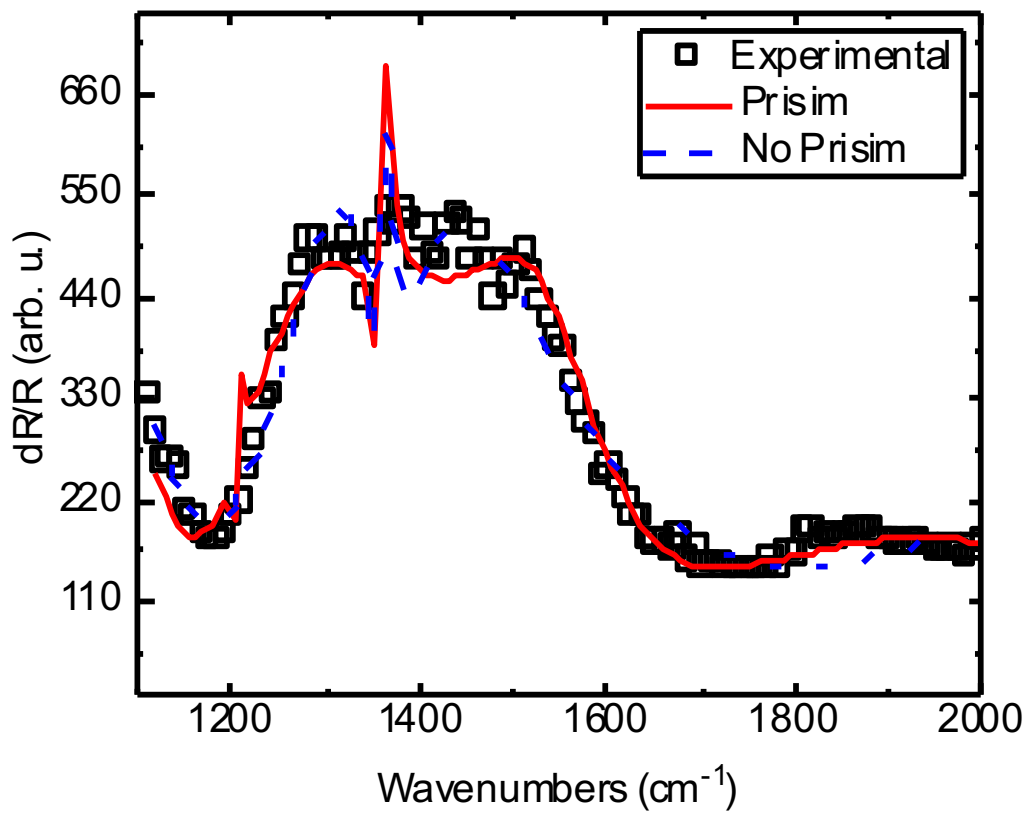

*Figure S12: The two fits with and without a prism on top of the measured signal, this figure is of a 500ps spectral cross-section.*

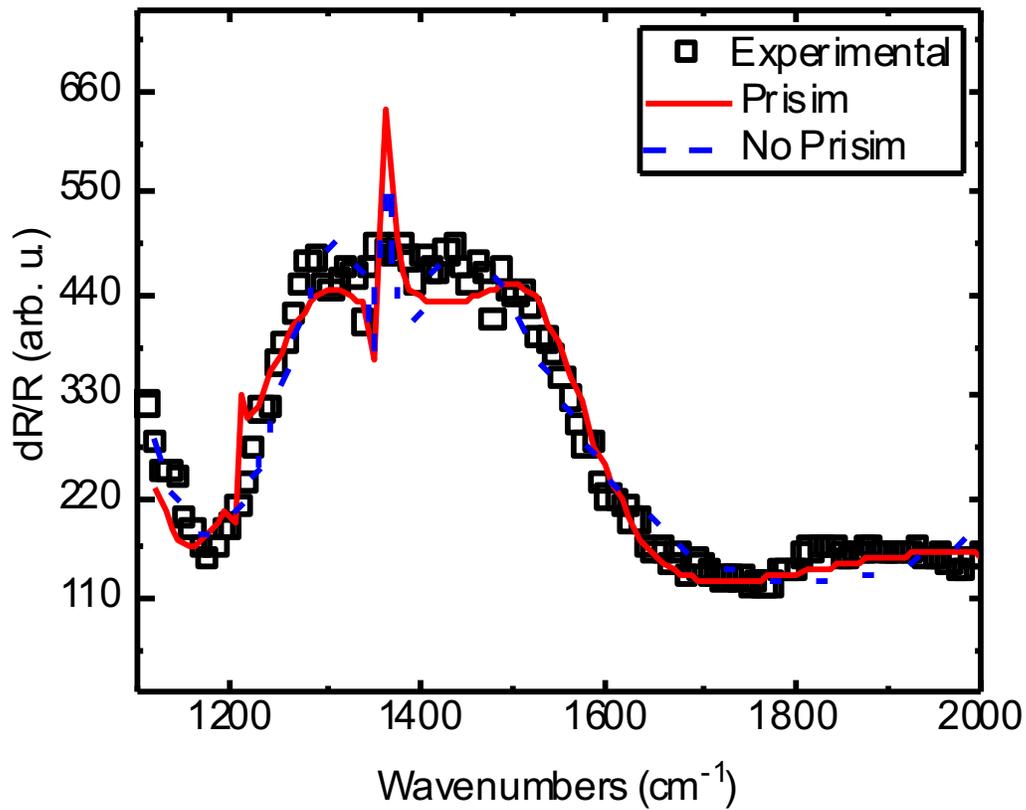

*Figure S13: The two fits with and without a prism on top of the measured signal, this figure is of a 800ps spectral cross-section.*

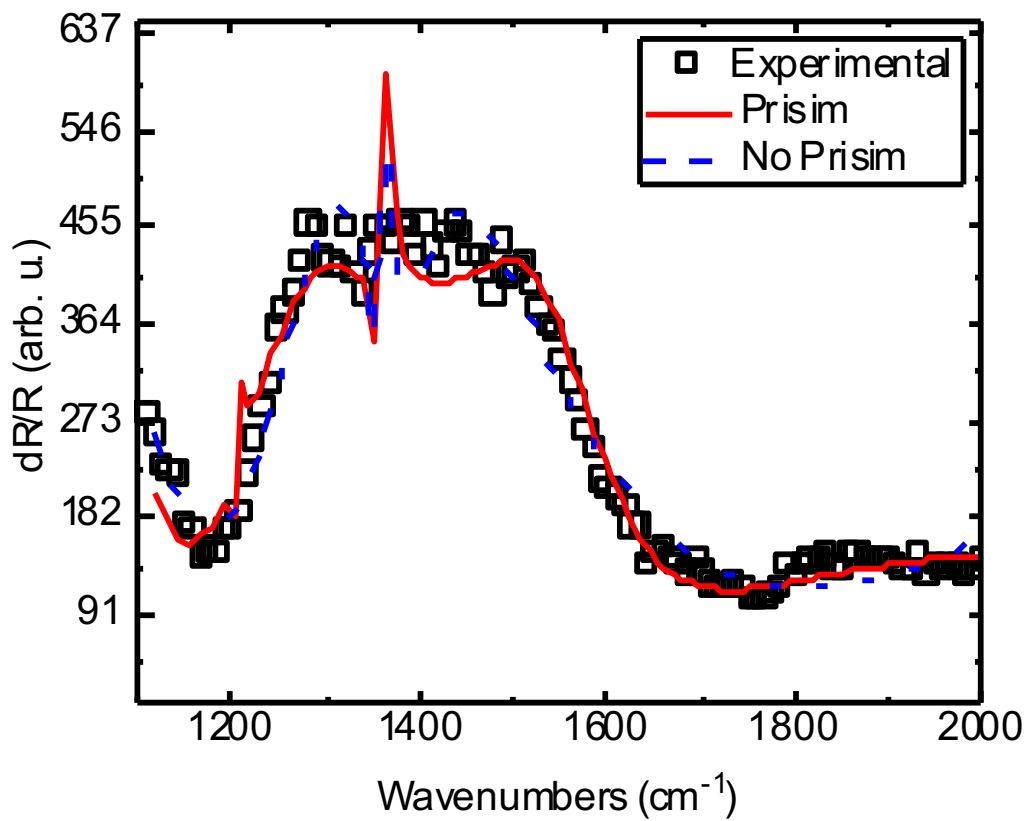

*Figure S3: The two fits with and without a prism on top of the measured signal, this figure is of a 1100ps spectral cross-section.*

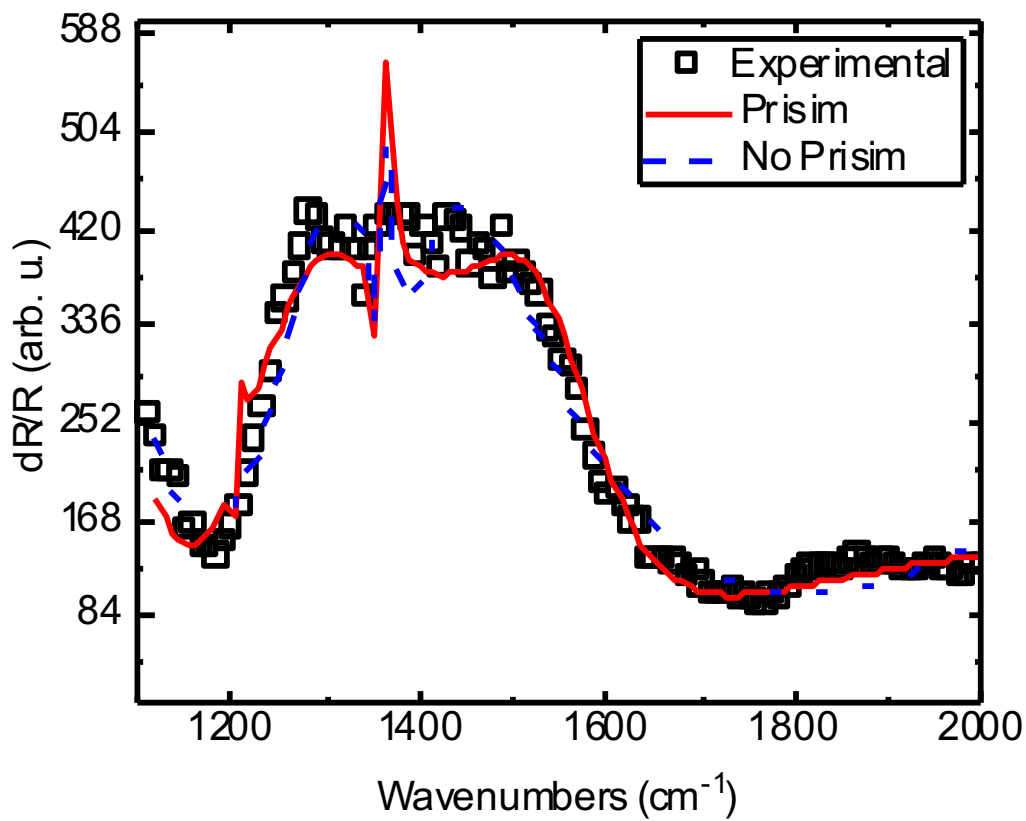

*Figure S4: The two fits with and without a prism on top of the measured signal, this figure is of a 2000ps spectral cross-section.*

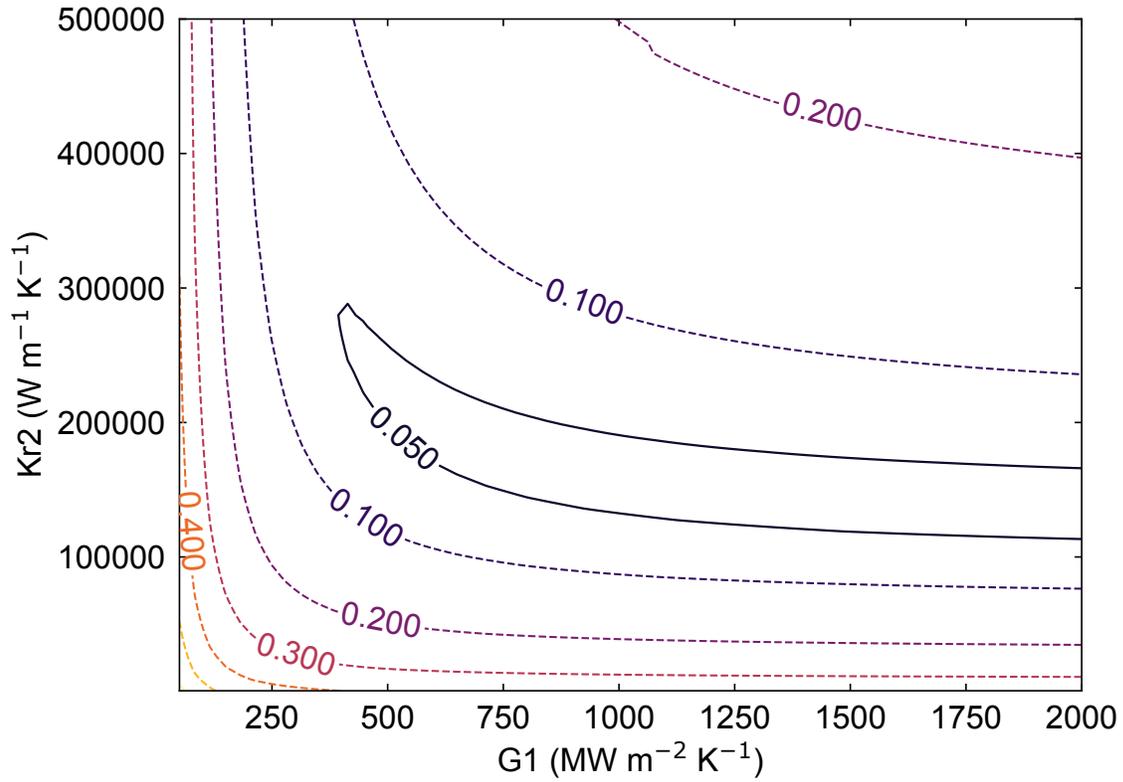

Figure S16: The Contour Uncertainty associated with the analytical model ffor approximating TBCs showing an lack on sensitivity to highTBCs >500.

|  | k [W m$^{-1}$ K$^{-1}$] | C [J m$^3$ K$^{-1}$] | h$_{AU-HBN}$ [W m$^{-2}$ K$^{-1}$] |
|---|---|---|---|
| Au | N/A | 2.5e6[20] | 1e9[FIT] |
| HBN | 3.5[8] | 1.28e6[1] | N/A |

Table S3: Values used for our POS model of Au-HBN. The optical phonon heat capacity for HBN was estimated assuming a Debye model for heat capacity and using a phonon dispersion found in literature[1]. The thermal conductivity of HBN is defined as the *geometric mean of the cross plane and in-plane thermal conductivities, $\sqrt{\kappa_r \kappa_z}$.*[8,21] *The polariton boundary conductance(PBC) was determined from fitting with an initial value of 1e9 W m$^{-2}$K$^{-1}$.*

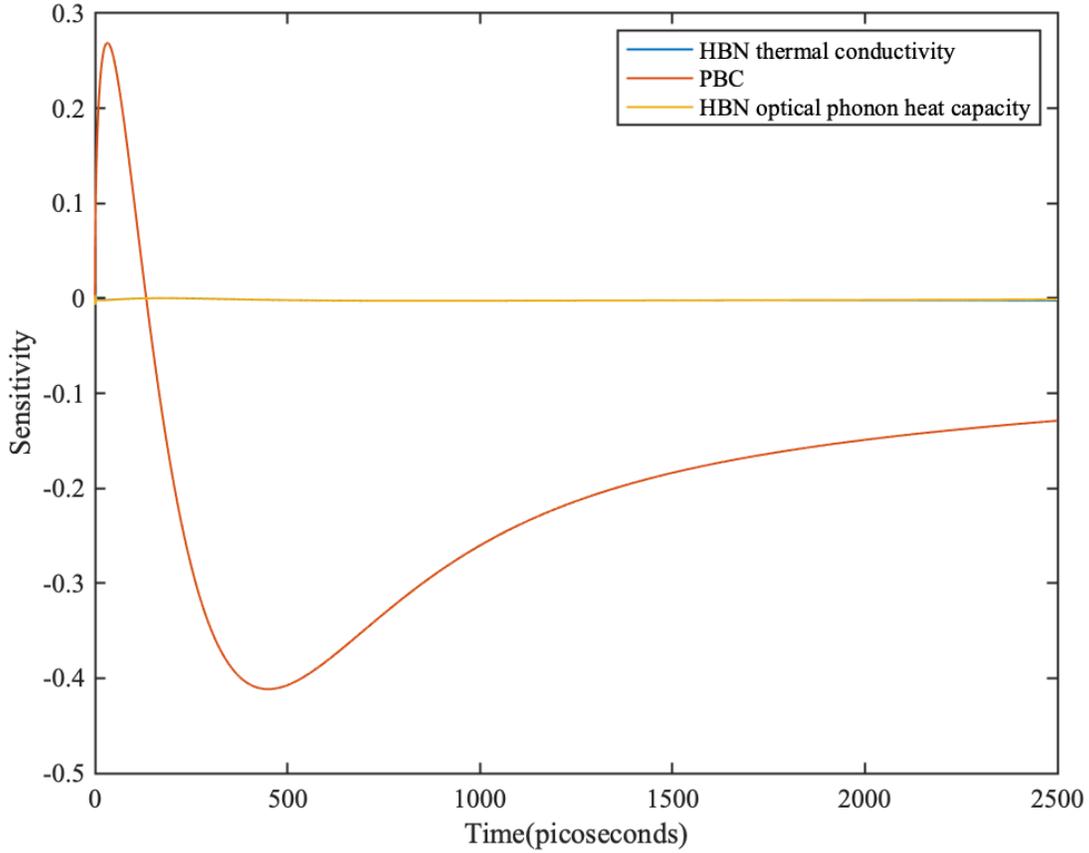

*Figure S17: Here we show sensitivities of the boron nitride temperature response to the thermal conductivity of boron nitride, polaritonic boundary conductance, and the heat capacity of boron nitride respectively calculated by methods of Schmidt et al.*[17] *We have high sensitivity to polaritonic boundary conductance at all times, whereas both the thermal conductivity and heat capacity of boron nitride only have slight sensitivities at early times.*

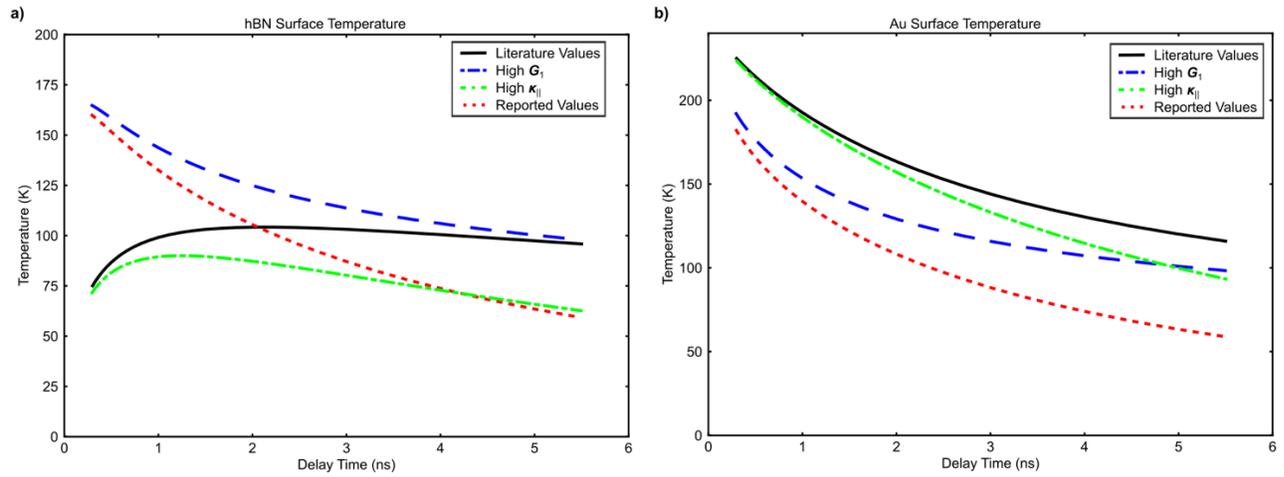

Figure S18: The calculated temperature vs time graphs predicted by the analytical model under experimental conditions. The maximum temperature rise in the hBN is predicted to be 80K under using the reported values from the main text. Without the high TBC ($G_1$), the curvature did not match the concavity in the measurements, and the high in-plane conductance ($\kappa_{||}$) serves to bring the general shape at long (>1ns) times

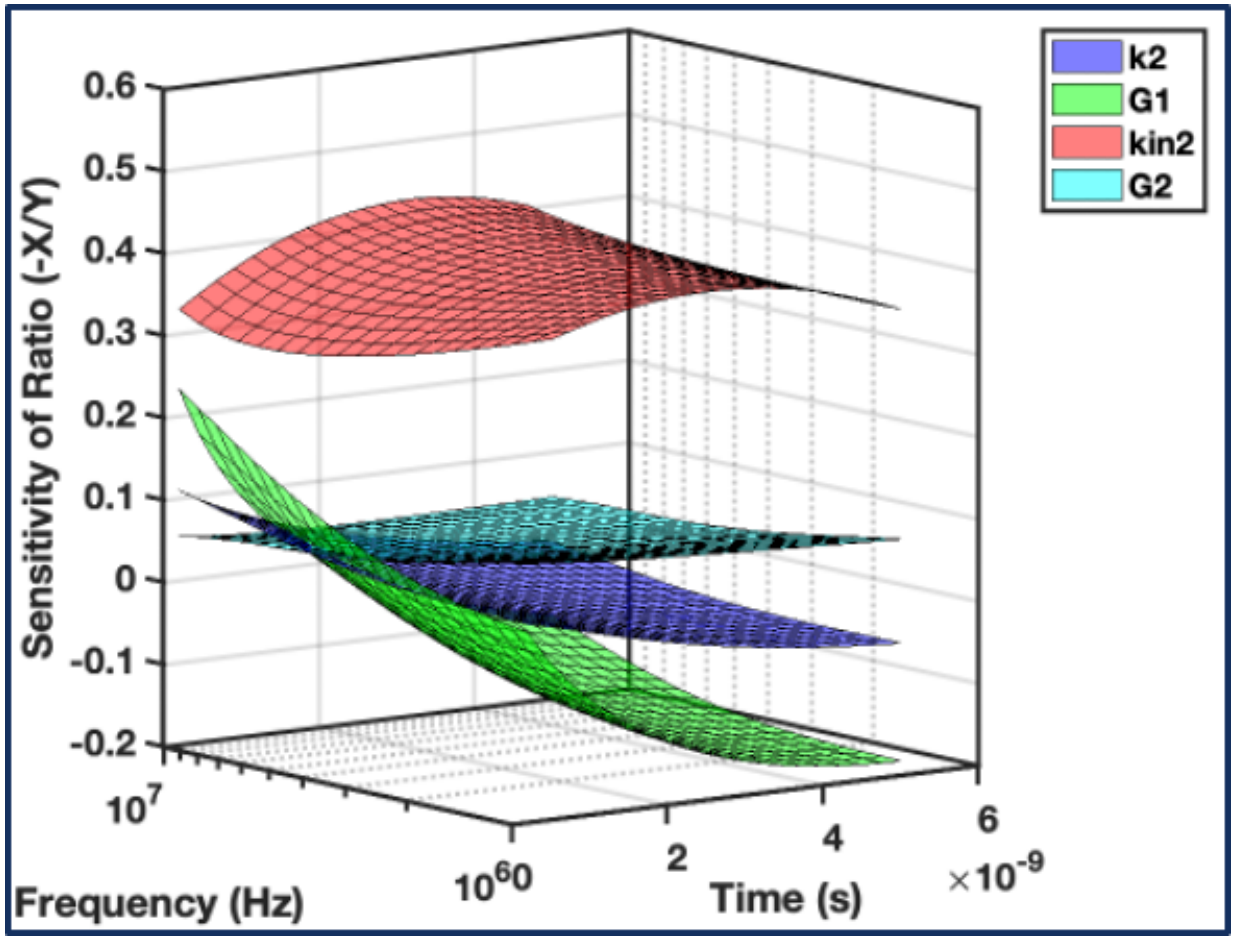

*Figure S19: A plot showing the sensitivity to thermal parameters of the stack measured within the time and frequency range of interest5.*

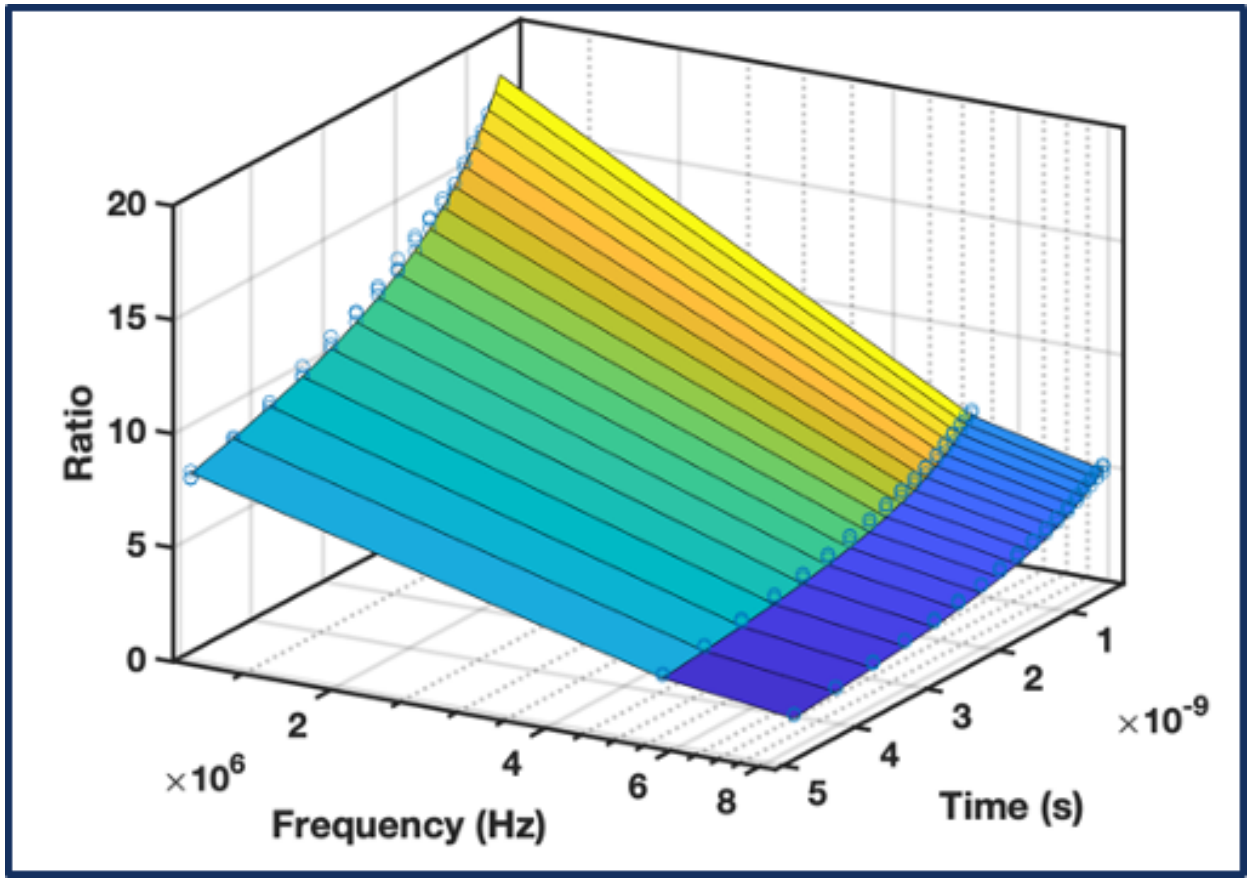

*Figure S20: The fitted contour of the TDTR measurments averaged over three scans per modulation frequency.*

|      | $k_{in}$ [W m$^{-1}$ K$^{-1}$] | C [J m$^3$ K$^{-1}$] | $h_{AU-HBN}$ [W m$^{-2}$ K$^{-1}$] | d[m] |
|------|---|---|---|---|
| Au   | 209[MEASURED] | 2.5e6[20] | 12.3e6±2.3e6[FIT] | 47e-9[MEASURED] |
| HBN  | 748±170.2[FIT] | 1.55e6[22] | N/A | 118e-9[MEASURED] |

*Table S4: Values used for our TDTR fit of Au-HBN thermal stack. The gold thermal conductivity was measured with a four-point probe station averaging over 10*

*measurements. Both hBN and Au thicknesses were measured via AFM. The thermal boundary conductance(TBC) was determined from fitting with an initial value of 20e6 W m$^{-2}$K$^{-1}$.*

E. **References**

1. Giles, A. J. *et al.* Ultralow-loss polaritons in isotopically pure boron nitride. *Nat Mater* **17**, 134–139 (2018).

2. Vuong, T. Q. P. *et al.* Isotope engineering of van der Waals interactions in hexagonal boron nitride. *Nature Materials 2017 17:2* **17**, 152–158 (2017).

3. Liu, S. *et al.* Single Crystal Growth of Millimeter-Sized Monoisotopic Hexagonal Boron Nitride. *Chemistry of Materials* **30**, 6222–6225 (2018).

4. Wilson, R. B. *et al.* Thermoreflectance of metal transducers for optical pump-probe studies of thermal properties. *Optics Express, Vol. 20, Issue 27, pp. 28829-28838* **20**, 28829–28838 (2012).

5. Hao, J. & Zhou, L. Electromagnetic wave scatterings by anisotropic metamaterials: Generalized 4×4 transfer-matrix method. *Phys Rev B Condens Matter Mater Phys* **77**, (2008).

6. Olmon, R. L. *et al.* Optical dielectric function of gold. *Phys Rev B Condens Matter Mater Phys* **86**, (2012).

7. Cahill, D. G. Analysis of heat flow in layered structures for time-domain thermoreflectance. *Review of Scientific Instruments* **75**, 5119–5122 (2004).

8. Jiang, P., Qian, X., Yang, R. & Lindsay, L. Anisotropic thermal transport in bulk hexagonal boron nitride. *Phys Rev Mater* **2**, (2018).

9. Chen, J. K., Tzou, D. Y. & Beraun, J. E. A semiclassical two-temperature model for ultrafast laser heating. *Int J Heat Mass Transf* **49**, 307–316 (2006).

10. Qiu, T. Q. & Tient, C. L. Femtosecond laser heating of multi-layer metals-I. Analysis Femtosecond laser heating of multi-layer metals-l. Analysis. *Article in International Journal of Heat and Mass Transfer* **37**, 994 (1994).

11. Hohlfeld, J. *et al.* Electron and lattice dynamics following optical excitation of metals. *Chem Phys* **251**, 237–258 (2000).

12. Hopkins, P. E., Norris, P. M. & Stevens, R. J. Influence of inelastic scattering at metal-dielectric interfaces. *J Heat Transfer* **130**, (2008).

13. Wang, W. & Cahill, D. G. Limits to Thermal Transport in Nanoscale Metal Bilayers due to Weak Electron-Phonon Coupling in Au and Cu. (2012) doi:10.1103/PhysRevLett.109.175503.

14. Wang, L., Cheaito, R., Braun, J. L., Giri, A. & Hopkins, P. E. Thermal conductivity measurements of non-metals via combined time- and frequency-domain thermoreflectance without a metal film transducer. *Review of Scientific Instruments* **87**, 094902 (2016).

15. Smits, F. M. Measurement the of Sheet Resistivities Four-Point Probe. *The Bell System Technical Journal* **37**, 711–718 (1958).

16. Kittel, C. *Introduction to Solid State Physics*. (Jhon Wiley & Sons, 1996).

17. Schmidt, A. J., Cheaito, R. & Chiesa, M. A frequency-domain thermoreflectance method for the characterization of thermal properties. *Rev. Sci. Instrum* **80**, 94901 (2009).

18. Francoeur, M. & Pinar Mengüç, M. Role of fluctuational electrodynamics in near-field radiative heat transfer. *J Quant Spectrosc Radiat Transf* **109**, 280–293 (2008).

19. Francoeur, M., Pinar Mengüç, M. & Vaillon, R. Solution of near-field thermal radiation in one-dimensional layered media using dyadic Green's functions and the scattering matrix method. *J Quant Spectrosc Radiat Transf* **110**, 2002–2018 (2009).

20. Tomko, J. A., Kumar, S., Sundararaman, R. & Hopkins, P. E. Temperature dependent electron-phonon coupling of Au resolved via lattice dynamics measured with sub-picosecond infrared pulses. *J Appl Phys* **129**, (2021).

21. Braun, J. L., Olson, D. H., Gaskins, J. T. & Hopkins, P. E. A steady-state thermoreflectance method to measure thermal conductivity. *Review of Scientific Instruments* **90**, 024905 (2019).

22. Yuan, C. *et al.* Modulating the thermal conductivity in hexagonal boron nitride via controlled boron isotope concentration. *Communications Physics 2019 2:1* **2**, 1–8 (2019).